\documentclass[fleqn,twoside]{article}%
\usepackage{amsmath}
\usepackage{amsfonts}
\usepackage{amssymb}
\usepackage{graphicx}%
\usepackage{caption}
\def\be{\begin{equation}}

\def\ee{\end{equation}}
\def\bes{\begin{equation}\begin{split}&}
\def\es{\end{split}}
\def\bi{\bibitem}

\begin{document}
\title{Probing early universe with a generalized action}
\author{Ranajit Mandal$^1$, Dalia Saha $^2$, Mohosin Alam $^3$, Abhik Kumar Sanyal $^4$}

\maketitle
\noindent
\begin{center}
${^1}$ Dept. of Physics, Rammohan College, Kolkata, West Bengal, India - 700009.\\
\noindent
$^{2, 4}$ Dept. of Physics, Jangipur College, Murshidabad, West Bengal, India - 742213\\
\noindent
${^3}$ Dept. of Physics, Saidpur U. N. H. S., Murshidabad, West Bengal, India - 742225. \\

\end{center}

\footnotetext[1]{
\noindent Electronic address:\\
$^1$ranajitmandalphys@gmail.com\\
$^2$daliasahamandal1983@gmail.com\\
$^3$alammohosin@gmail.com\\
$^4$sanyal\_ ak@yahoo.com\\}
\noindent

\begin{abstract}
Possibly, the most general action in the background of isotropic and homogeneous space-time has been considered to study the quantum evolution of the early universe, apart from a cosmological constant. The hermiticity of the effective Hamiltonian operator in the presence of curvature squared terms suggests unitary time evolution of the quantum states, assuring conservation of probability. The oscillatory behaviour of the semi-classical wavefunction around a de-Sitter solution signals that the theory is classically allowed, and the universe enters an inflationary regime just after Planck's era. In view of a hierarchy of Hubble flow parameters, and using a redefined effective potential, the complicated classical field equations in the presence of several coupling parameters, reduce to standard general-relativistic equations with a single scalar field. As a result, inflation has been studied without considering any additional flow parameters. Inflationary parameters lie very much within the presently available Planck's data, and the model admits graceful exit from inflation.

\end{abstract}
\noindent
\section{Introduction:}

In quantum mechanics, `unitarity' is a restriction on the allowed evolution of quantum systems that ensures the sum of probabilities of all possible outcomes of an event is always normalized to `one'. So the operator which describes the evolution of a physical system in time, must be unitary. Likewise, the S-matrix, that describes how the physical system changes in scattering process, must also be a unitary operator implying optical theorem \footnote{Optical theorem is a consequence of the conservation of probability. In wave scattering theory, it relates the forward scattering amplitude to the extinction cross-section of the scatterer.}. Therefore, the time evolution in quantum theory must be formulated as a unitary transformation generated by the Hamiltonian. Such formulation is possible in `Non-relativistic Quantum Theory' where Hamiltonian is the total energy, and also in `Special Theory of Relativity' where Hamiltonian is the time-component of the four-momenta. Nevertheless, problems appear in `General Theory of Relativity' (GTR), since in GTR neither energy nor the momenta are local quantities, rather they are defined globally and only for suitable asymptotic behaviour. Further, the fact that gauge-invariant divergences make GTR non-renormalizable, is quite familiar by now. Non-renormalizable theories are acceptable as description of low energy physics, but these theories have intrinsic mass-scale at which the effective low energy theories break down. For GTR it is the Planck's scale ($M_P = 10^{19}~\mathrm{GeV}$). The Non-renormalizability of GTR indicates that one should opt for a new physics at Planck's scale. Consequently, if Einstein-Hilbert action for GTR is modified in a manner such that principle candidates are the contracted quadratic products of the curvature tensor, then fourth derivative terms appear which lead to a suitable graviton propagator that behaves like $k^{-4}$ for large momenta, and the resulting action \cite{Stelle}

\be \label{1.1} A = \int \left[\alpha R + \beta_1 R^2 + \beta_2\Big( {1\over 3} R^2 - R_{\mu\nu}R^{\mu\nu}\Big)\right]\sqrt{-g} d^4x,\ee
is renormalizable. Once renormalization is established for pure gravity, the inclusion of other renormalized fields through coupling, does not pose any further problem. However, it was also simultaneously realized that such fourth order theories lead to ghosts when expanded in the perturbative series about the linearized theory, since the linearized energy of the five massive spin-2 excitations is negative definite \cite{Stelle}. This posses major obstacles to their physical interpretation, since it destroys the `unitarity', and thus appears to preclude the acceptability of the above action \eqref{1.1} as a physical theory. On the contrary, there are counter arguments which indicate that the problem with unitarity appearing in perturbative expansion might be misleading. First of all, for $N$ number of matter fields, the theory is unitary in $1\over N$ expansion, as $N \rightarrow \infty$ \cite{Tomb1}. Secondly, the standard non-perturbative `Osterwalder-Schrader' construction \footnote{it satisfies the condition that correlation functions on Euclidean space-time have to be equivalent to the correlation functions of a Wightman Quantum Field Theory on Minkowski space-time, i.e. it assures that the Wick rotation is well defined isomorphism of `Quantum Field Theory' on Minkowski as well as on Euclidean space-time.} resulted in a Hilbert space with positive norm, which proves the unitarity of fourth order gravitational action \eqref{1.1}. Particularly, the action is found to be asymptotically free \cite{Tomb2, Tomb3, Tset}. Asymptotic freedom is the property that causes interactions between particles to become arbitrarily weak at arbitrarily large energy scales, corresponding to arbitrarily small length scale. Thus asymptotic free theories are non-perturbatively renormalizable. Thirdly, it is pointed out that the presence of a massive spin-2 ghost in the bare propagator is inconclusive, since this excitation is unstable. It is shown that the physical S-matrix between in and out states containing only transverse, massless gravitons and physical massless matter fields are gauge independent, and the contribution of all gauge-variant poles to its intermediate states must cancel. The physical S-matrix should therefore, be unitary \cite{Anti, Tomb4}. Fourthly, up on quantization of the effective Hamiltonian corresponding to the conformal version of the above action \eqref{1.1}, no ghosts are found to the leading order, in a strong coupling expansion  \cite{Kaku}. Finally, no ghosts are seen at the classical level, in the zero total energy theorem, even without Einstein-Hilbert term of the above action \eqref{1.1} \cite{Boul}. Of course, notwithstanding the fact that with the advent of the theories like superstring, loop quantum gravity, supergravity etc., these higher order theories we are talking of are presently viewed simply as the low energy effective theories, the issue of unitarity of such effective actions is not precluded from further study. \\

One of the major difficulties in formulating a quantum gravity theory is that quantum gravitational effects only appear near the Planck length scale, $l_P \sim 10^{-35}$ m. Nonetheless, in the absence of a complete quantum theory of gravity, quantum cosmology renders certain physical insights near Planck's era, which is essentially the motivation of the present work. However, this requires Hamiltonian formulation of the action \eqref{1.1} at the first place,  and canonical quantization thereafter, which is of-course a non-trivial task due to the presence of higher order curvature invariant terms. Problem is simplified by and large, if one starts with the cosmological principle, i.e. treats the universe to be isotropic and homogeneous a-priori (while observable anisotropy appears due to scalar and metric perturbations), and considers Robertson-Walker metric described by,

\be \label{RW} ds^2 = -N^2dt^2+a(t)^2\left[\frac{dr^2}{1-kr^2}+r^2(d\theta^2+sin^2\theta\; d\phi^2)\right],\ee
where, $N$ is the lapse function. Note that the term $\big({1\over 3} R^2 - R_{\mu\nu}R^{\mu\nu}\big)\sqrt{-g}~d^4x$ appearing in the action \eqref{1.1} is a total derivative term, and thus does not contribute to the field equations. Further, a more general action should incorporate $R_{\alpha\beta\gamma\delta}R^{\alpha\beta\gamma\delta}$ term, which appears in the so-called Gauss-Bonnet combination, viz. $\mathcal{G}\sqrt{-g}~d^4x = (R^2 - 4R_{\alpha\beta}R^{\alpha\beta} + R_{\alpha\beta\gamma\delta}R^{\alpha\beta\gamma\delta})\sqrt{-g}~d^4x$. Although, this Gauss-Bonnet term kills the ghost, unfortunately it is again topologically invariant in $4$-dimensions. Nonetheless, both these terms may be incorporated in the action \eqref{1.1}, to obtain non-trivial contributions in the field equations, through suitable coupling terms, once a scalar field is taken into account. In that case, an action in its most general form, may be expressed as,

\be\label{A}
\begin{split}& A = \int\left[\alpha(\phi)R+\beta_1(\phi)R^2+\beta_2(\phi)\Big(R_{\mu\nu}^2-{1\over 3}R^2\Big)+\gamma(\phi)\mathcal{G}-\frac{1}{2}\phi_{,\mu}\phi^{,\mu}-V(\phi)\right] d^{4}x \sqrt{-g},\end{split} \ee
apart from a pure cosmological constant term. In the above, despite the required functional dependence of $\beta_2(\phi)$ and $\gamma(\phi)$, we have also chosen functional dependence of $\alpha(\phi)$ and $\beta_1(\phi)$, for generality, while $V(\phi)$ is an arbitrary potential. The presence of the term being coupled to the functional parameter $\beta_2(\phi)$ makes the present action even more general than the one considered earlier \cite{RDMA}. Particularly, $\gamma(\phi) \mathcal{G}$ is called the Gauss-Bonnet-dilatonic coupling term, which arises naturally as the leading order of the $\alpha'$ expansion of heterotic superstring theory, where, $\alpha'$ is the inverse string tension \cite{S1, S2, S3, S4}. Further, the low energy limit of the string theory also gives rise to the dilatonic scalar field which is found to be coupled with various curvature invariant terms \cite{D1, D2}. Therefore, the leading quadratic correction gives rise to Gauss-Bonnet term with a dilatonic coupling \cite{BD}. It is important to mention that the dilatonic coupled Gauss-Bonnet term plays an important role at the late-stage of cosmic evolution, exhibiting accelerated expansion after a long Friedmann-like deceleration, in the matter dominated era \cite{I1, I2}. For a clarification, we remind that conformally invariant theory of gravity is specified by the conformal Weyl squared term, for which the Lagrangian density reads as $\mathcal{L} = \alpha R - {1\over m^2}C_{\alpha\beta\gamma\delta} C^{\alpha\beta\gamma\delta}$, $m$ being the mass scale at which such correction becomes relevant. Nonetheless, it is worthy to mention that Weyl square term vanishes in $4$-dimensional isotropic metric. Note that the expression for Weyl squared term is $C^2 = {1\over 3}R^2 - 2R_{\alpha\beta}R^{\alpha\beta} + R_{\alpha\beta\gamma\delta}R^{\alpha\beta\gamma\delta} = \mathcal{G} -2({1\over 3}R^2 - R_{\alpha\beta}R^{\alpha\beta})$. Since, in action \eqref{A} the two terms on the right hand side appear with two different coupling parameters, so it is even more general than incorporating Weyl squared term.\\

In the following section, we write the general field equations and their counterparts in the background of isotropic and homogeneous Robertson-Walker minisuperspace \eqref{RW}. Classical de-Sitter solutions are thereby explored. In section 3, we perform canonical analysis as a mere prologue to canonical quantization scheme, followed by an appropriate semiclassical approximation. In section 4, inflation under slow-roll approximation is performed and the results are compared with recently released data sets from Planck's collaborators \cite{pd1, pd2}. Finally we conclude in section 5.

\section{Action, field equations and classical solutions:}

The field equation corresponding to the action \eqref{1.1} is found under the standard metric variation as \cite{fld},

\be\label{gfield}\begin{split} &2\bigg(\alpha(\phi)G_{\mu\nu}+\Box\alpha(\phi) g_{\mu\nu} -\nabla_{\mu}\nabla_{\nu}\alpha(\phi)\bigg) +4\bigg(\beta_1(\phi)- {\beta_2(\phi)\over 3} \bigg) RR_{\mu\nu}-4g_{\mu\nu} \Box\Big{[}\bigg(\beta_1(\phi)- {\beta_2(\phi)\over 3} \bigg)R\Big]-4\nabla_{\mu}\nabla_{\nu} \\& \Big[\bigg(\beta_1(\phi)- {\beta_2(\phi)\over 3} \bigg) R\Big] -g_{\mu\nu} \bigg(\beta_1(\phi)- {\beta_2(\phi)\over 3} \bigg)R^2 +2\beta_2(\phi)\big(g_{\mu\nu} R_{\alpha \beta} R^{\alpha\beta} -4 R_{\mu\alpha}R_{\nu}^{\alpha}\big)
-4g_{\mu\nu}\nabla_{\alpha}\nabla_{\beta}\beta_2(\phi)R^{\alpha\beta} \\& +8\nabla_{\alpha}\nabla_{\nu}\beta_2(\phi)R_{\mu}^{\alpha}-4\Box\beta_2(\phi) R_{\mu\nu} +2\gamma(\phi)H_{\mu\nu} +8\big(\gamma''\nabla^{\rho}\phi\nabla^{\sigma}\phi+\gamma'\nabla^{\rho}\nabla^{\sigma}\phi\big)P_{\mu\rho\nu\sigma}
-T_{\mu\nu}=0,\end{split}\ee
where, $G_{\mu\nu}=R_{\mu\nu}-{1\over 2}g_{\mu\nu}R$ and $T_{\mu\nu}=\nabla_{\mu}\phi\nabla_{\nu}\phi-{1\over 2}g_{\mu\nu} \nabla_{\lambda}\phi\nabla^{\lambda} \phi-g_{\mu\nu}V(\phi)$ are the Einstein tensor and the energy-momentum tensor, respectively. Further, $H_{\mu\nu}=2\{RR_{\mu\nu}-2R_{\mu\rho}R^{\rho}_{\nu}-2R_{\mu\rho\nu\sigma}R^{\rho\sigma}+R_{\mu\rho\sigma\lambda}R^{\sigma\rho\lambda}_\nu-{1\over 2}g_{\mu\nu}\mathcal{G} \}$ and $P_{\mu\nu\rho\sigma}=R_{\mu\nu\rho\sigma}+2g_{\mu[\sigma} R_{\rho]\nu}+2g_{\nu[\rho} R_{\sigma]\mu}+Rg_{\mu[\rho} g_{\sigma]\nu}$. Explicit form of equation \eqref{gfield} together with  the $\phi$ variation equation may be written as,

\be \label{phivar}\begin{split} &\left(\alpha G_{\mu\nu} + \Box\alpha g_{\mu\nu} - \alpha_{;\mu;\nu}\right) + 2\Big(\beta_1- {\beta_2\over 3} \Big)\Big(RR_{\mu\nu} - {1\over 4}g_{\mu\nu}R^2\Big) -2\left[\Box{\Big\{ \Big(\beta_1- {\beta_2\over 3} \Big) R\Big\}} g_{\mu\nu} +\Big\{\Big(\beta_1- {\beta_2\over 3} \Big) R\Big\}_{;\mu;\nu}\right]\\& +\beta_2g_{\mu\nu} R_{\alpha \beta} R^{\alpha\beta} -4\beta_2 R_{\mu\alpha} R_{\nu}^{\alpha} -2g_{\mu\nu}\nabla_{\alpha}\nabla_{\beta}\beta_2R^{\alpha\beta}+ 4\nabla_{\alpha} \nabla_{\nu}\beta_2 R_{\mu}^{\alpha}-2\Box\beta_2 R_{\mu\nu} \\&+ 2\gamma \Big[RR_{\mu\nu} - 2 R_{\mu\rho}R^{\rho}_{\nu} - 2R_{\mu\rho\nu\sigma}R^{\rho\sigma} + R_{\mu\rho\sigma\lambda}R_\nu^{\sigma\rho\lambda} - {1\over 2}g_{\mu\nu}\mathcal{G}\Big]\\& +4\left(\gamma'' \phi^{;\rho}\phi^{;\sigma}+\gamma' \phi^{;\rho;\sigma}\right) \big[R_{\mu\rho\nu\sigma} + 2g_{\mu[\sigma}R_{\rho]\nu} + 2g_{\nu[\rho}R_{\sigma]\mu} + R g_{\mu[\rho}g_{\sigma]\nu} \big] = {T_{\mu\nu}\over 2};\\&
\mathrm{and},\\&
\Box\phi-\alpha' R-\beta_1' R^2-\beta_2'\big( R_{\mu\nu}^2-{1\over 3}R^2 \big)-\gamma' \mathcal{G}-V'=0, \end{split}\ee
respectively, where prime denotes derivative with respect to $\phi$. In the homogeneous and isotropic Robertson-Walker metric \eqref{RW}, the Ricci scalar reads as,

\be \label{R} {R}=\frac{6}{N^2}\left(\frac{\ddot{a}}{a}+\frac{\dot{a}^{2}}{a^2}+N^2\frac{k}{a^{2}}-\frac{\dot a\dot N}{aN}\right).\ee
Not all the components of Einstein's equations are independent. It therefore suffices to write the two independent components of Einstein's field equations, viz., the $(^0_0)$ equation and the $\phi$ variation equation, under standard gauge choice $N=1$, in terms of the scale factor $a$ as,

    \be \label{00} \begin{split} -\frac{6\alpha}{a^2}\bigg{(}\dot a^2+ k\bigg{)}-\frac{6\alpha' \dot a\dot\phi}{a} &-36\beta_1 \bigg{(}\frac{2\dot a\dddot a}{a^2}-\frac{\ddot a^2}{a^2} +\frac{2\dot a^2 \ddot a}{a^3}-\frac{3\dot a^4}{a^4}-{2k\dot a^2\over a^4}+{k^2\over a^4}\bigg{)}-72\beta_1'\dot\phi\bigg{(}\frac{\dot a\ddot a}{a^2}+\frac{\dot a^3}{a^3}+{k\dot a\over a^3}\bigg{)}\\&+{6\beta_2'\dot\phi}\bigg({2\dot a^3\over a^3}+{3k\dot a\over a^3} \bigg)-24\gamma'\dot\phi\bigg{(}{\dot a^3\over a^3}+{k\dot a\over a^3}\bigg{)}+\bigg{(}\frac{\dot\phi^2}{2}+V \bigg{)}=0\end{split} \ee
and
 \be \label{phivariation} \begin{split} -6\alpha'\bigg{(}a^2\ddot a+a\dot a^2+ka\bigg{)}& -36\beta_1'\bigg{(}a\ddot a^2+2\dot a^2\ddot a+ \frac{\dot a^4}{a}+{k^2\over a}+{2k\dot a^2\over a}+2k\ddot a\bigg{)} +12\beta_2'\bigg({\dot a^2\ddot a}+{k\ddot a} \bigg)\\&+3a^2\dot a \dot\phi -24\gamma'\bigg{(}\dot a^2\ddot a+k\ddot a\bigg{)} +a^3\bigg{(}\ddot\phi+V'\bigg{)}=0 .\end{split}\ee
It is not difficult to see that the above field equations also admit the following de-Sitter solution in the spatially flat space ($k=0$),

\be\label{aphi} \begin{split} a = &a_0e^{\lambda t}; ~~~ \phi= \phi_0e^{-\lambda t},~~\mathrm{under~ the~ condition},\\&
\alpha(\phi) = {\alpha_0\over \phi};~~ V(\phi) = \frac{1}{2} \lambda^2 \phi^2;~~\mathrm{and}~~ \beta_2-2\Big(6 \beta_1 +  \gamma\Big) = {1\over \lambda^2}\left({\alpha_0\over \phi} +{\phi^2\over 24}\right),\end{split}\ee
 where, $a_0$, $\phi_0$, $\alpha_0$ and $\lambda$ are arbitrary constants while $\beta_i(\phi)$ $\gamma(\phi)$ are arbitrary functions of $\phi$ related as above after setting the constant of integration to zero without any loss of generality. Since we shall encounter operator ordering ambiguities during canonical quantization, therefore we remove the arbitrariness on $\beta_i$ and $\gamma$, following a simple assumption viz,

\be\label{const} \beta_1 = \frac{\alpha_0}{12\lambda^2\phi} = {\beta_{01}\over \phi};\hspace{0.2 cm}\beta_2 = \frac{2\alpha_0}{\lambda^2\phi} = {\beta_{02}\over \phi};\hspace{0.2 cm}\gamma = -\frac{\phi^2}{48\lambda^2} = {\gamma_0\phi^2}\hspace{0.2 cm} \mathrm{where},\hspace{0.2 cm}\beta_{01} = {\alpha_0\over 12 \lambda^2};\hspace{0.2 cm}\beta_{02} = {2\alpha_0\over \lambda^2};~~\gamma_0 = -{1\over 48\lambda^2}.\ee
The above forms of $\alpha$ \eqref{aphi}, $\beta_i$ and $\gamma$ \eqref{const} will be used during canonical quantization (for operator ordering, in particular), semiclassical approximation, as well as in the slow roll approximation.

\section{Canonical formulation:}

The field equations corresponding to the action \eqref{1.1}, associated with present higher-order curvature invariant terms are of fourth order, and hence canonical formulation requires one additional degree of freedom. Ostrogradski's technique \cite{Ostro} towards canonical formulation of higher order theories does not work for the singular Lagrangian (for which the determinant of the Hessian vanishes) under consideration, at least due to the presence of the Lapse function $N$, which essentially is a Lagrange multiplier. It is therefore required to follow Dirac’s algorithm of constrained analysis \cite{Dirac1, Dirac2}. In Dirac's formalism, for treating higher-order theory of gravity, it is customary to assume $\delta {h_{ij}}|_{\partial V}= 0 = \delta {K_{ij}}|_{\partial V}$ at the boundary, where, $h_{ij}$ is the induced three metric, and $K_{ij}$ is the extrinsic curvature tensor. However, Modified Horowitz' Formalism (MHF) \cite{We1, We2, We3, We4, We5, We6, We7, We8, We9, We10, We11, We12, We13} bypasses the constrained analysis. In MHF $\delta {h_{ij}}|_{\partial V}= 0 = \delta R|_{\partial V}$ at the boundary, and the action is required to be supplemented by appropriate boundary terms. There exists other techniques too, and the Hamiltonians obtained following different techniques are canonically equivalent \cite{Buch}. Nevertheless, equivalence at the quantum level requires canonical transformation in the quantum domain, and since at the classical level canonical transformations are highly non-linear, so the different quantum descriptions so obtained are likely to be inequivalent \cite{Deru}. In this connection it has recently been established \cite{We12} that MHF and the Dirac constraint analysis towards canonical formulation of higher-order theory of gravity lead to the same Hamiltonian, and therefore to identical quantum description, provided the action is first expressed in terms of $h_{ij}$, and Dirac algorithm is initiated only after taking care of the divergent terms appearing in the action. We shall here follow the MHF, while Dirac's technique is exhibited in the appendix. As mentioned, in order to expatiate MHF, the action \eqref{A} should be supplemented by appropriate boundary terms, viz.,

\begin{center}
\be\label{Scf}
\begin{split}A=\int  \bigg{[}{\alpha(\phi)R}+\beta_1(\phi)R^2+&\beta_2(\phi)\Big(R_{\mu\nu}^2-{1\over 3}R^2\Big) +\gamma(\phi)\mathcal{G}-\frac{1}{2}\phi_{,\mu}\phi^{,\mu}-V(\phi)\bigg{]}\sqrt{-g}~d^{4}x\\& +\alpha(\phi)\Sigma_{R}+\beta_1(\phi) \Sigma_{R^2} +\beta_2(\phi)\Sigma_{{[}R_{\mu\nu}^2-{1\over 3}R^2{]}}+\gamma(\phi)\Sigma_{\mathcal{G}}.\end{split} \ee
\end{center}
In the above, the supplementary boundary terms $\Sigma_{R}=2\oint_{\partial\nu}K\sqrt{h}d^3x; \Sigma_{R^2}=\Sigma_{R_1^2}+\Sigma_{R_2^2}=4\oint_{\partial\nu}[ {^3R}+ ({^4R}-{^3R})]RK\sqrt{h}d^3x=4\oint_{\partial\nu}RK\sqrt{h}d^3x; \Sigma_{\mathcal{G}}=4\oint_{\partial\nu}\big{(}2G_{ij}K^{ij}+ \frac{\textsf{K}}{3}\big{)}\sqrt{h}d^3x$ and $\Sigma_{{[}R_{\mu\nu}^2-{1\over 3}R^2{]}}$ are the Gibbons-Hawking-York term and its subsequent modifications \cite{GHY, GHY1, GHY2}, while $\textsf{K}=K^2-3KK^{ij}K_{ij}+2K^{ij}K_{ik}K^k_j$, $K$ being the trace of extrinsic curvature tensor $K_{ij}$. Let us briefly enunciate our programme. First, we express the action (\ref{Scf}) in terms of the basic variable $h_{ij} = a^2\delta_{ij} = z\delta_{ij}$, where, $a^2 = z$, and remove the divergent terms under integration by parts. Next, we shall introduce an auxiliary variable, remove divergent terms again, and finally translate the auxiliary variable to the other basic variable, viz. $K_{ij} = -\frac{\dot h_{ij}}{2N}=-\frac{a\dot a}{N}\delta_{ij}=\frac{\dot z}{2N}\delta_{ij}$, to obtain the phase-space structure. So we first write the action (\ref{Scf}) in terms of $h_{ij} = a^2\delta_{ij} = z\delta_{ij}$ using the form of the Ricci scalar \eqref{R} as,

\begin{center}\be \label{Sc1}\begin{split} A=\int &\bigg{[}{3\alpha(\phi)\sqrt{z}\bigg{(}\frac{\ddot z}{N}-\frac{\dot z\dot N}{N^2}+2kN\bigg{)}}+\frac{9\beta_1(\phi)}{\sqrt z}\bigg{(}\frac{\ddot z^2}{N^3}-\frac{2\dot z \ddot z \dot N}{N^4}+\frac{\dot z^2\dot N^2}{N^5}-\frac{4k\dot z\dot N}{N^2}+\frac{4k\ddot z}{N}+4k^2N \bigg{)}\\&-\beta_2(\phi)\bigg[{3\dot z^2\ddot z\over 2N^3z^{3\over 2}}-{3\dot z^3\dot N\over 2N^4z^{3\over 2}}-{{3\dot z^4\over 4N^3z^{5\over 2}}}+{6k\ddot z\over N\sqrt z}-{6k\dot z \dot N\over N^2\sqrt z}-{3k\dot z^2\over Nz^{3\over 2}}\bigg]\\&+\frac{3\gamma(\phi)}{N\sqrt z}\bigg{(}\frac{\dot z^2\ddot z}{N^2z}-\frac{\dot z^4}{2N^2z^2}-\frac{\dot z^3\dot N}{N^3z}+4k\ddot z-\frac{2k\dot z^2}{z}-\frac{4k\dot z\dot N}{N} \bigg{)}+z^{\frac{3}{2}}\bigg{(}\frac{1}{2N}\dot{\phi}^2-NV(\phi)\bigg{)}\bigg{]}dt\\& + \alpha(\phi)\Sigma_{R}
+\beta_1(\phi)\Sigma_{R^2}+\beta_2(\phi)\Sigma_{{[}R_{\mu\nu}^2-{1\over 3}R^2{]}}+\gamma(\phi)\Sigma_{\mathcal{G}}.\end{split}\ee\end{center}
In the above, $\Sigma_{R}=-\frac{3\sqrt z \dot z}{N}$, $\Sigma_{R_1^2}=-\frac{36 k \dot z}{N\sqrt z}$, $\Sigma_{R_2^2}=-\frac{18\dot z}{N^3\sqrt z}\textbf{(}\ddot z-\frac{\dot z\dot N}{N}\textbf{)}$, $\Sigma_{{[}R_{\mu\nu}^2-{1\over 3}R^2{]}}=\big{(}{\dot z^3\over 2N^3z^{3\over 2}}+{6k\dot z\over N\sqrt z} \big{)}$ and $\Sigma_{\mathcal{G}}=-\frac{\dot z}{N\sqrt z}\textbf{(}\frac{\dot z^2}{N^2 z}+12k\textbf{)}$ supplementary surface terms known as Gibbons-Hawking-York (GHY) and its subsequent modified versions in the isotropic and homogeneous Robertson-Walker metric \eqref{RW}. It is important to mention that unlike GTR, here the lapse function appears in the action with its time derivative, behaving like a true variable. Despite such uncanny situation, one can still bypass Dirac's algorithm, as we demonstrate underneath. First, under integrating the above action \eqref{Sc1} by parts, the counter terms $\Sigma_R$, $\Sigma_{R_1^2}$, $\Sigma_{{[}R_{\mu\nu}^2- {1\over 3}R^2{]}}$ and $\Sigma_{\mathcal{G}}$ get cancelled and the above action (\ref{Sc1}) reads as,

\be \label{Sc2}\begin{split} A&=\int \bigg{[}{\bigg{(}-\frac{3\alpha'\dot\phi\dot z\sqrt z}{N}-\frac{3\alpha\dot z^2}{2N\sqrt z}+6kN\alpha\sqrt z\bigg{)}}+\frac{9\beta_1}{\sqrt z}\bigg{(}\frac{\ddot z^2}{N^3}-\frac{2\dot z \ddot z \dot N}{N^4}+\frac{\dot z^2\dot N^2}{N^5}+\frac{2k{\dot z}^2}{Nz}+4k^2N \bigg{)}\\&-\frac{36k\beta'_1 \dot z \dot\phi}{N\sqrt z}+\beta'_2\dot\phi\bigg{(}{\dot z^3\over 2N^3z^{3\over 2}}+{6k\dot z\over N\sqrt z} \bigg{)}-\frac{\gamma'\dot z \dot\phi}{N\sqrt z}\bigg{(}\frac{\dot z^2}{N^2z}+12k \bigg{)}+z^{\frac{3}{2}}\bigg{(}\frac{\dot{\phi}^2}{2N}-NV\bigg{)}\bigg{]}dt +\beta_1(\phi)\Sigma_{R_2^2}. \end{split}\ee
At this stage we introduce an auxiliary variable,

\be\label{Q} Q=\frac{\partial A}{\partial\ddot z}=\frac{18\beta_1}{N^3\sqrt z}\bigg{(}\ddot z-\frac{\dot N\dot z}{N}\bigg{)},\ee
judicially into the action (\ref{Sc2}), so that it takes the following form,

\be \label{Sc3} \begin{split} A&=\int \bigg{[} {\bigg{(}}-\frac{3\alpha'\dot\phi \dot z\sqrt z}{N}-\frac{3\alpha \dot z^2}{2N\sqrt z}+6kN\alpha \sqrt z \bigg{)}+\bigg{(}Q\ddot z-\frac{N^3\sqrt z Q^2}{36\beta_1}-\frac{\dot N\dot z Q}{N}\bigg{)}+\frac{9\beta_1}{\sqrt{z}}\bigg(\frac{2k\dot z^2}{N z}+ 4k^2N\bigg{)}\\& -\frac{36k\beta'_1 \dot \phi \dot z}{N\sqrt z}+\beta'_2\dot\phi\bigg{(}{\dot z^3\over 2N^3z^{3\over 2}}+{6k\dot z\over N\sqrt z} \bigg{)}-\frac{\gamma'\dot z \dot\phi}{N\sqrt z}\bigg{(}\frac{\dot z^2}{N^2z}+12k \bigg{)}+z^{\frac{3}{2}}\bigg{(}\frac{\dot{\phi}^2}{2N}-NV\bigg{)}\bigg{]}dt +\beta_1(\phi)\Sigma_{R_2^2}.\end{split}\ee
Now, integrating by parts yet again, the last of the surface terms gets cancelled with the total derivative term and the action (\ref{Sc3}) can finally be expressed as,

\be \label{Sc4} \begin{split}A=\int \bigg{[} {\bigg{(}}&-\frac{3\alpha'\dot\phi \dot z\sqrt z}{N}-\frac{3\alpha \dot z^2}{2N\sqrt z}+6kN\alpha \sqrt z \bigg{)}-\bigg(\dot Q\dot z+\frac{N^3\sqrt z Q^2}{36\beta_1}+\frac{\dot N\dot z Q}{N}\bigg)+\frac{9\beta_1}{\sqrt{z}}\bigg(\frac{2k\dot z^2}{N z}+ 4k^2N\bigg{)}\\& -\frac{36k\beta'_1 \dot \phi \dot z}{N\sqrt z}+\beta'_2\dot\phi\bigg{(}{\dot z^3\over 2N^3z^{3\over 2}}+{6k\dot z\over N\sqrt z} \bigg{)} -\frac{\gamma'\dot z \dot\phi}{N\sqrt z}\bigg{(}\frac{\dot z^2}{N^2z}+12k \bigg{)}+z^{\frac{3}{2}}\bigg{(}\frac{\dot{\phi}^2}{2N}-NV\bigg{)}\bigg{]}dt.\end{split}\ee
Therefore, the canonical momenta are
\be\label{Mom}\begin{split}&
p_Q =-\dot z, \\&
p_z =-\frac{3\alpha'\dot\phi\sqrt z}{N}-\frac{3\alpha \dot z}{N\sqrt z}-\dot Q-\frac{\dot N Q}{N}+\frac{36k\beta_1 \dot z}{Nz^{\frac{3}{2}}}-\frac{36k\beta'_1 \dot\phi}{N\sqrt z}+{\beta'_2\dot\phi}\bigg({3\dot z^2\over 2N^3z^{3\over 2}}+{6k \over N\sqrt z}\bigg)-\frac{3\gamma'\dot\phi}{N\sqrt z}\bigg{(}\frac{\dot z^2}{N^2 z}+4k\bigg{)},\\&
p_{\phi}=-\frac{3\alpha'\dot z\sqrt z}{N}-\frac{36k\beta'_1 \dot z}{N\sqrt z}+\beta'_2\bigg{(}{\dot z^3\over 2N^3z^{3\over 2}}+{6k\dot z\over N\sqrt z} \bigg{)}-\frac{\gamma'\dot z}{N\sqrt z}\bigg{(}\frac{\dot z^2}{N^2 z}+12k\bigg{)}+\frac{z^{\frac{3}{2}} \dot\phi}{N}, \\&
p_N = -\frac{Q\dot z}{N}.
\end{split}\ee
The action \eqref{Sc4} still contains time derivative of the Lapse function $N$ and not all the momenta are invertible, implying degeneracy of the Lagrangian. One can bypass Dirac's constraint analysis as claimed earlier, up on finding the following relationship, in view of the definition of momenta \eqref{Mom},

\be\label{PP} p_Q p_z=\frac{3\alpha' \dot z \dot\phi \sqrt z}{N}+\frac{3\alpha \dot z^2}{N\sqrt z}+\dot z\dot Q+\frac{\dot N\dot z Q}{N}-\frac{36k\beta_1 \dot z^2}{Nz^{\frac{3}{2}}}+\frac{36k\beta'_1 \dot z \dot\phi}{N\sqrt z}-3\beta'_2\dot\phi\Big({\dot z^3\over 2N^3z^{3\over 2}}+{2k\dot z \over N\sqrt z}\Big)+\frac{3\gamma'\dot z  \dot\phi}{N\sqrt z}\bigg{(}\frac{\dot z^2}{N^2 z}+4k\bigg{)},\ee
and using the above relation \eqref{PP}, and the definitions of momenta \eqref{Mom}, to obtain the phase space structure of the Hamiltonian constraint equation as,

\be \label{Sc5} \begin{split} H_c&=-p_Q p_z +\frac{N^3Q^2\sqrt z}{36\beta_1}+\frac{Np_{\phi}^2}{2z^{\frac{3}{2}}}-\frac{3\alpha' p_Q p_{\phi}}{z}-\frac{36k\beta'_1{p_Q} {p_{\phi}}}{z^2}+{N\beta'_2p_{\phi}\over z^{3\over 2}}\bigg({p_Q^3\over 2N^3z^{3\over 2}}+ {6kp_Q\over N\sqrt z}\bigg)-{\gamma'p_Q p_{\phi}\over z^2}\bigg{(}{p_Q^2\over N^2z}+{12k}\bigg{)}\\&+3\alpha \bigg{(}\frac{{p_Q^2}}{2N\sqrt z}-2kN\sqrt z\bigg{)}+\frac{9\alpha'^2 p_Q^2}{2N\sqrt z}-\frac{18k\beta_1}{\sqrt z}\bigg{(}\frac{{p_Q^2}}{Nz}+2kN\bigg{)}+\frac{648k^2\beta_1'^2{p_Q^2}}{Nz^{\frac{5}{2}}}-{3\alpha'\beta'_2p_Q\over z}\bigg({p_Q^3\over 2N^3 z^{\frac{3}{2}}}+{6kp_Q\over N\sqrt z} \bigg)\\&+\frac{108k\alpha'\beta'_1 p_Q^2}{Nz^{\frac{3}{2}}}+{3\alpha'\gamma'p_Q^2\over Nz^{3\over 2}}\bigg{(}{p_Q^2\over N^2z}+{12k}\bigg{)}-{36k\beta'_1 \beta'_2p_Q\over z^2}\bigg({p_Q^3\over 2N^3z^{3\over 2}}+{6kp_Q\over N\sqrt z} \bigg)+{36k\beta_1'\gamma'p_Q^2\over Nz^{5\over 2}}\bigg{(}{p_Q^2\over N^2z}+{12k}\bigg{)}\\&+{N\beta_2'^2\over 2z^{3\over 2}}\bigg{(}{p_Q^3\over 2N^3z^{3\over 2}}+{6kp_Q\over N\sqrt z} \bigg{)}^2-{\beta'_2\gamma' p_Q\over z^2}\bigg{(}{p_Q^3\over 2N^3z^{3\over 2}}+{6kp_Q\over N\sqrt z} \bigg{)}\bigg( {p_Q^2\over N^2z}+12k\bigg)+{\gamma'^2p_Q^2\over 2Nz^{5\over 2}}\bigg{(}{p_Q^2\over N^2z}+{12k}\bigg{)}^2+Nz^{\frac{3}{2}}V=0. \end{split}\ee
The problem with the above Hamiltonian is that, firstly it does not exhibit diffeomorphic invariance $H = N\mathcal{H}$, and second, the momenta $p_Q$ appears upto sixth degree. This uncanny situation may easily be alleviated, upon replacing the auxiliary variable ($Q$) by the basic variable. In fact, under the following canonical transformations $Q=\frac{p_x}{N}$ and $p_Q=-Nx$, the phase-space structure of the Hamiltonian can be expressed in terms of basic variables as,

\be \label{Hc} \begin{split} H_c&= N\bigg{[}xp_z+ \frac{\sqrt z p_x^2}{36\beta_1}+\frac{p_{\phi}^2}{2z^{\frac{3}{2}}}+\frac{3\alpha' x p_{\phi}}{z}+\frac{36k\beta_1'{x} {p_{\phi}}}{z^2}-{\beta_2' p_{\phi}\over z^{3\over 2}}\bigg({x^3\over 2z^{3\over 2}}+{6kx\over \sqrt z} \bigg)+{\gamma'x p_{\phi}\over z^2}\bigg({x^2\over z}+{12k}\bigg)+\frac{9\alpha'^2 x^2}{2\sqrt z}\\& +3\alpha \bigg{(}\frac{x^2}{2\sqrt z}-2k\sqrt z\bigg{)}+\frac{108k\alpha'\beta_1' x^2}{z^{\frac{3}{2}}}-{3\alpha'\beta_2' x\over z}\bigg({x^3\over 2z^{3\over 2}}+{6kx\over \sqrt z} \bigg)+{3\alpha'\gamma'x^2\over z^{3\over 2}}\bigg({x^2\over z}+{12k}\bigg)+\frac{648k^2\beta_1'^2{x}^2}{z^{\frac{5}{2}}}\\&-\frac{18k\beta_1}{\sqrt z}\bigg{(}\frac{x^2}{z}+2k\bigg{)}-{36k\beta_1'\beta_2' x\over z^2}\bigg({x^3\over 2z^{3\over 2}}+{6kx\over \sqrt z} \bigg)+{36k\beta_1'\gamma'x^2\over z^{5\over 2}}\bigg({x^2\over z}+{12k}\bigg)+{\beta_2'^2\over 2z^{3\over 2}}\bigg({x^3\over 2z^{3\over 2}}+{6k x\over \sqrt z} \bigg)^2\\& -{\beta_2'\gamma' x\over z^2}\bigg({x^3\over 2z^{3\over 2}}+{6k x\over \sqrt z} \bigg)\bigg({x^2\over z}+12k \bigg)+{\gamma'^2x^2\over 2z^{5\over 2}}\bigg({x^2\over z}+{12k}\bigg)^2+z^{\frac{3}{2}}V\bigg{]}=N\mathcal{H}=0. \end{split}\ee
In the process, both the problems are alleviated, i.e. diffeomorphic invariance is established in one hand, and the momenta also appear only with second degree. It is now also possible to express the action (\ref{Sc3}) in the canonical form with respect to the basic variables as,

\be \label{CF} A=\int\bigg{(}\dot z p_z+\dot x p_x+\dot\phi p_{\phi}-N\mathcal{H}\bigg{)}dt d^3x=\int\bigg{(}\dot{h}_{ij}\pi^{ij}+\dot{K}_{ij}{\Pi}^{ij}+\dot\phi p_{\phi}-N\mathcal{H}\bigg{)}dt d^3x,\ee
where $\pi^{ij}$ and $\Pi^{ij}$ are momenta canonically conjugate to $h_{ij}$ and $K_{ij}$ respectively.

\subsection{Canonical quantization:}

The quantum counterpart of the Hamiltonian (\ref{Hc}) under standard canonical quantization reads as,
\be \label{qh} \begin{split} \frac{i\hbar}{\sqrt z}\frac{\partial\Psi}{\partial z}&=\bigg{[}-\frac{\hbar^2}{36\beta_1 x}\bigg{(}\frac{\partial^2}{\partial x^2} +\frac{n}{x}\frac{\partial}{\partial x}\bigg{)}-\frac{\hbar^2}{2xz^2}\frac{\partial^2}{\partial \phi^2}+\frac{36k\hat{\beta_1'}\hat{p}_\phi}{z^{\frac{5}{2}}}   -\frac{\hat{\beta_2'}\hat{p}_\phi}{xz^2}\bigg({x^3\over 2z^{3\over 2}}+{6kx\over \sqrt z} \bigg)+\frac{\hat{\gamma'}\hat{p}_\phi}{z^\frac{5}{2}}\bigg({x^2\over z}+{12k}\bigg)\\&+\frac{3\hat{\alpha}'\hat{p}_\phi}{z^{\frac{3}{2}}}+\frac{108k\hat{\alpha}'\hat{\beta_1'}x}{z^2}-{3\hat{\alpha}'\hat{\beta_2'}\over z^{3\over 2}}\bigg({x^3\over 2z^{3\over 2}}+{6kx\over \sqrt z} \bigg)+\frac{3\hat{\alpha}'\hat{\gamma}'x}{z^2}\bigg({x^2\over z}+{12k}\bigg)-{36k\hat{\beta_1'}\hat{\beta_2'}\over z^{5\over 2}}\bigg({x^3\over 2z^{3\over 2}}+{6kx\over \sqrt z} \bigg)\\&+\frac{648k^2\hat{\beta_1'}^2x}{z^3}+{36k\hat{\beta_1'}\hat{\gamma}'x\over z^3}\bigg({x^2\over z}+{12k}\bigg)+{\hat{\beta_2'}^2\over 2z^2x}\bigg({x^3\over 2z^{3\over 2}}+{6k x\over \sqrt z} \bigg)^2-{\hat{\beta_2'}\hat{\gamma}'\over z^{5\over 2}}\bigg({x^3\over 2z^{3\over 2}}+{6k x\over \sqrt z} \bigg)\bigg({x^2\over z}+12k \bigg)\\&+\frac{9\hat{\alpha}'^2x }{2z}+{\hat{\gamma}'^2x\over 2z^3}\bigg({x^2\over z}+{12k}\bigg)^2+\bigg{(}\frac{3\alpha x}{2z}-\frac{6k\alpha}{x} -\frac{18kx\beta_1}{z^2} -\frac{36k^2\beta_1}{xz}+\frac{Vz}{x}\bigg{)}\bigg{]}\Psi, \end{split}\ee
where, $n$ is the operator ordering index which removes some but not all of the operator ordering ambiguities appearing between $\{\hat{x}, \hat{p}_x\}$. Now, in order to remove additional ambiguities in connection with the pairs $\{\hat{\alpha}',\hat{p}_{\phi}\}$, $\{\hat{\beta_1'},\hat{p}_{\phi}\}$, $\{\hat{\beta_2'},\hat{p}_{\phi}\}$, $\{\hat{\gamma}',\hat{p}_{\phi}\}$ etc., we need to know the functional dependence of the coupling parameters. At this stage, we use the functional forms of $\alpha(\phi)$, $\beta_1(\phi), \beta_2(\phi)$ and $\gamma(\phi)$ vide \eqref{aphi} and \eqref{const}, in view of the classical de-Sitter solution. Thus, performing Weyl symmetric ordering carefully, equation (\ref{qh}) takes the following form,

\begin{center}
    \be \label{qh1} \begin{split} \frac{i\hbar}{\sqrt z}\frac{\partial\Psi}{\partial z}=\bigg{[}&-\frac{\hbar^2\phi}{36\beta_{01} x}\bigg{(}\frac{\partial^2}{\partial x^2} +\frac{n}{x}\frac{\partial}{\partial x}\bigg{)}-\frac{\hbar^2}{2xz^2}\frac{\partial^2}{\partial \phi^2}+\frac{3i\hbar \alpha_0} {z^{\frac{3}{2}}} \bigg{(}\frac{1}{\phi^2}\frac{\partial}{\partial \phi}-\frac{1}{\phi^3}\bigg{)}-\frac{i\hbar\beta_{02} x^2}{2 z^{\frac{7}{2}}} \bigg{(}\frac{1}{\phi^2}\frac{\partial}{\partial \phi}-\frac{1}{\phi^3}\bigg{)}\\& -\frac{i\hbar\gamma_0 x^2}{z^{\frac{7}{2}}} \bigg{(}2\phi \frac{\partial}{\partial\phi}+1\bigg{)}+\frac{9\alpha_0^2 x}{2 \phi^4 z}-\frac{3\alpha_0\beta_{02} x^3}{2 z^3 \phi^4}-\frac{6\alpha_0\gamma_0 x^3}{\phi z^3}+\frac{\beta_{02}^2 x^5}{8 \phi^4 z^5}+\frac{\beta_{02}\gamma_0 x^5}{\phi z^5}+\frac{2 \gamma_{0}^2\phi^2 x^5}{z^5}\\&+\frac{3 \alpha_{0} x}{2 \phi z}+\frac{\lambda^2 \phi^2 z}{2 x}\bigg{]}\Psi. \end{split}\ee
\end{center}
Now, under a change of variable $\sigma^2=z^3$, the above modified Wheeler-de-Witt equation, takes the look of Schr\"{o}dinger equation, viz.,

\begin{center}
    \be \label{qh2} \begin{split} {i\hbar}\frac{\partial\Psi}{\partial \sigma} =&\bigg{[}-\frac{\hbar^2\phi}{54\beta_{01} x}\bigg{(}\frac{\partial^2}{\partial x^2} +\frac{n}{x}\frac{\partial}{\partial x}\bigg{)}-\frac{\hbar^2} {3x\sigma^{\frac{4}{3}}}\frac{\partial^2}{\partial \phi^2}+\frac{2i\hbar \alpha_0} {\sigma} \bigg{(}\frac{1}{\phi^2}\frac{\partial}{\partial \phi}-\frac{1}{\phi^3}\bigg{)} -\frac{i\hbar \beta_{02} x^2} {3\sigma^\frac{7}{3}} \bigg{(}\frac{1}{\phi^2}\frac{\partial}{\partial \phi}-\frac{1}{\phi^3}\bigg{)}\\&-\frac{2i\hbar\gamma_0x^2}{3\sigma^{\frac{7}{3}}} \bigg{(}2\phi \frac{\partial}{\partial\phi}+1\bigg{)}+\hat V_e\bigg{]}\Psi=\hat{H_e}\Psi. \end{split}\ee
\end{center}
In the above Schr\"{o}dinger-like equation, the effective potential $V_e$ is given by,
\begin{center}
    \be\begin{split} \hat V_e = V_e &=\frac{3\alpha_0^2x}{\sigma^{\frac{2}{3}}\phi^2}+{\beta_{02}^2x^5\over 12\sigma^{10\over 3}\phi^4}-{\alpha_0\beta_{02}x^3\over \sigma^2\phi^4}-{4\alpha_0\gamma_0 x^3\over \sigma^2\phi}+{2\beta_{02}\gamma_0 x^5\over 3\sigma^{10\over 3}\phi}+{4\gamma_0^2x^5\phi^2\over 3\sigma^{10\over 3}}+{\alpha_0 x\over \sigma^{2\over 3}\phi}+{\lambda^2\sigma^{2\over 3}\phi^2\over 3x},\end{split} \ee
\end{center}
and, $\sigma=z^{\frac{3}{2}}=a^3$ plays the role of internal time parameter. It is quite important to mention that, since time itself acts as a dynamical variable in the theory of gravity, the Hamiltonian appears as a constraint, and upon quantization for any state $\Psi$, $\hat{H}|\Psi\rangle = 0$, time disappears. Thus GTR confronts with standard probabilistic interpretation, and one is not supposed to ask what happened earlier, since time collapses. However, despite the fact that the proper volume of the universe itself is a dynamical variable, it acts as an internal time parameter in the quantum description of higher-order theory, and standard quantum mechanical probabilistic interpretation holds, as we explore in the following subsection.

\subsection{Hermiticity of $\hat{H}_e$ and probabilistic interpretation:}

We initiated our discussion in connection with unitarity, which is the fundamental requirement of a viable quantum theory. In fact, unitarity and consistency are synonym, and the hermiticity of a time independent Hamiltonian leads to unitary time evolution, which assures conservation of probability. Further, in quantum scattering theory, hermiticity is necessary both for reciprocity and unitarity. Thus, the requirement of `hermiticity' is necessary and sufficient condition for the unitary time evolution. In the following, we demonstrate that the effective Hamiltonian so obtained, is a hermitian operator. For this purpose, we split the effective Hamiltonian $\hat H_e$ obtained in (\ref{qh2}), and express it as,
\begin{center}
\be\label{He}\begin{split} &\hat H_e = \hat H_1 + \hat H_2 + \hat H_3 + \hat V_e,\end{split}\ee
\end{center}
where,
    \begin{eqnarray}
    \hat H_1 &=& -\frac{\hbar^2 \phi}{54\beta_{01} x}\bigg{(}\frac{\partial^2}{\partial x^2} +\frac{n}{x}\frac{\partial}{\partial x}\bigg{)} \\
       \hat H_2 &=& -\frac{\hbar^2}{3x\sigma^{\frac{4}{3}}}\frac{\partial^2}{\partial \phi^2} \\
       \hat H_3 &=& {i\hbar } \bigg{(}\frac{2\alpha_0}{\sigma\phi^2} -{4\gamma_0 x^2\phi\over 3\sigma^{7\over 3}}-{\beta_{02}x^2\over 3\sigma^{7\over 3}\phi^2}\bigg{)} \frac{\partial}{\partial \phi} +\frac{i\hbar}{\sigma} \bigg{(}-{2\alpha_0\over \phi^3}-{2\gamma_0x^2\over 3\sigma^{4\over 3}}+{\beta_{02}x^2\over 3\sigma^{4\over 3}\phi^3}\bigg{)}\\
       \hat V_e &=& V_e.
     \end{eqnarray}
Now, let us consider the first term,
\begin{center}
    \be \int \big{(}\hat H_1\Psi\big{)}^*\Psi dx= -\frac{\hbar^2 \phi}{54\beta_{01}}\int\bigg{(}\frac{1}{x}\frac{\partial^2\Psi^*}{\partial x^2} +\frac{n}{x^2}\frac{\partial\Psi^*}{\partial x}\bigg{)}\Psi dx=-\frac{\hbar^2 \phi}{54\beta_{01}}\int\bigg{(}\frac{\Psi}{x}\frac{\partial^2\Psi^*}{\partial x^2} +\frac{n\Psi}{x^2}\frac{\partial\Psi^*}{\partial x}\bigg{)} dx.\ee
\end{center}
Under integration by parts twice and dropping the first term due to fall-of condition, we obtain,
\begin{center}
    \be\label{H1} \int \big{(}\hat H_1\Psi\big{)}^*\Psi dx=-\frac{\hbar^2\phi}{54\beta_{01}}\int \Psi^*\bigg{[}\frac{1}{x}\frac{\partial^2\Psi}{\partial x^2} -\frac{n+2}{x^2}\frac{\partial\Psi}{\partial x}+\frac{2(n+1)}{x^3}\Psi\bigg{]}dx. \ee
\end{center}
Up on choosing the operator ordering index, $n=-1$, (\ref{H1}) turns out to be,
\begin{center}
    \be\label{H1.1} \int \big{(}\hat H_1\Psi\big{)}^*\Psi dx = -\frac{\hbar^2 \phi}{54\beta_{01}}\int \Psi^*\bigg{[}\frac{1}{x}\frac{\partial^2\Psi}{\partial x^2} -\frac{1}{x^2}\frac{\partial\Psi}{\partial x}\bigg{]}dx=\int \Psi^*\hat H_1\Psi dx. \ee
\end{center}
Thus, $\widehat H_1$ is hermitian, for a particular choice of operator ordering parameter, $n=-1$. Now since it is trivial to prove that $\hat H_2$ is hermitian, let us consider the third term, viz. $\hat H_3$,

\be \int (\hat H_3 \Psi)^*\Psi d\phi= -{i\hbar }\int \bigg{(}\frac{2\alpha_0}{\sigma\phi^2} -{4\gamma_0 x^2\phi\over 3\sigma^{7\over 3}}-{\beta_{02}x^2\over 3\sigma^{7\over 3}\phi^2}\bigg{)} \frac{\partial\Psi^*}{\partial \phi}\Psi d\phi -\frac{i\hbar}{\sigma}\int \bigg{(}-{2\alpha_0\over \phi^3}-{2\gamma_0x^2\over 3\sigma^{4\over 3}}+{\beta_{02}x^2\over 3\sigma^{4\over 3}\phi^3}\bigg{)} \Psi^*\Psi d\phi.\ee
Under integration by parts and dropping the integrated out terms due to fall-of condition, we obtain,
\begin{center}
    \be \int (\hat H_3 \Psi)^*\Psi d\phi= {i\hbar } \int\Psi^* \bigg{(}\frac{2\alpha_0}{\sigma\phi^2} -{4\gamma_0 x^2\phi\over 3\sigma^{7\over 3}}-{\beta_{02}x^2\over 3\sigma^{7\over 3}\phi^2}\bigg{)} \frac{\partial\Psi}{\partial \phi} d\phi +\frac{i\hbar}{\sigma}\int \bigg{(}-{2\alpha_0\over \phi^3}-{2\gamma_0x^2\over 3\sigma^{4\over 3}}+{\beta_{02}x^2\over 3\sigma^{4\over 3}\phi^3}\bigg{)} \Psi^*\Psi d\phi=\int \Psi^*\widehat H_3 \Psi d\phi,\ee
\end{center}
indicating $\hat H_3$ is hermitian too. Thus, the effective Hamiltonian $\hat H_e$ turns out to be a hermitian operator. The hermiticity of $\hat H_e$ now allows one to write the continuity equation in its standard form as,
\begin{center}
    \be \frac{\partial\rho}{\partial\sigma}+\nabla\textbf{.} \textbf{J}=0, \ee
\end{center}
in the following manner. This requires to find $\frac{\partial\rho}{\partial\sigma}$, where, $\rho = \Psi^*\Psi$, is the probability density and $\mathbf{J}$ is the current density. A little algebra leads to the following equation,
\begin{center}
    \be\begin{split} \frac{\partial\rho}{\partial\sigma} =  -\frac{\partial}{\partial x}\bigg{[}\frac{i\hbar \phi}{54\beta_{01} x}\big{(}\Psi\Psi^*_{,x}-\Psi^*\Psi_{,x} \big{)}\bigg{]}&-\frac{\partial}{\partial\phi}\bigg{[}\frac{i\hbar}{3x\sigma^{\frac{4}{3}}} \big{(}\Psi \Psi^*_{,\phi}-\Psi^*\Psi_{,\phi} \big{)}-\bigg{(}\frac{2\alpha_0}{\sigma\phi^2} -{4\gamma_0 x^2\phi\over 3\sigma^{7\over 3}}-{\beta_{02}x^2\over 3\sigma^{7\over 3}\phi^2}\bigg{)}\Psi^*\Psi\bigg{]} \\&-i\hbar\phi\frac{(n+1)}{54\beta_{01}x^2}\big{(}\Psi\Psi^*_{,x}-\Psi^*\Psi_{,x} \big{)}. \end{split} \ee
\end{center}
Clearly, the continuity equation can be expressed, only under the choice $n = -1$ as,
\begin{center}
    \be \frac{\partial\rho}{\partial\sigma}+\frac{\partial{J}_x}{\partial x}+\frac{\partial{J}_\phi}{\partial \phi}=0, \ee
\end{center}
where the current density $\textbf{J}=({J}_x, {J}_{\phi}, 0)$, and,
\begin{eqnarray}
  {J}_x &=& \frac{i\hbar\phi}{54\beta_{01} x}\big{(}\Psi\Psi^*_{,x}-\Psi^*\Psi_{,x} \big{)}, \\
  {J}_{\phi} &=&  \frac{i\hbar}{3x\sigma^{\frac{4}{3}}} \big{(}\Psi \Psi^*_{,\phi}-\Psi^*\Psi_{,\phi} \big{)}- \bigg{(}\frac{2\alpha_0}{\sigma\phi^2} -{4\gamma_0 x^2\phi\over 3\sigma^{7\over 3}}-{\beta_{02}x^2\over 3\sigma^{7\over 3}\phi^2}\bigg{)} \Psi^*\Psi.
\end{eqnarray}
Here, as already mentioned, the variable $\sigma$ plays the role of internal time parameter.

\subsection{Semiclassical approximation:}

Unitarity only proves the viability of a quantum equation in quantum domain. A quantum equation can only play an effective role in the physical world, if it admits an appropriate semiclassical approximation. Semiclassical approximation is essentially a method of finding an approximate wavefunction associated with a quantum equation. If the integrand in the exponent of the semiclassical wavefunction is imaginary, then the behaviour of the approximate wave function is oscillatory, and falls within the classical allowed region. Otherwise it is classically forbidden. Of-course, a quantum theory is justified, only when semiclassical approximation works, i.e. admits classical limit. Consequently, when the classical limit is admissible, most of the important physics are inherent in the classical action. A quantum theory therefore, may only be accepted as viable, if it admits and also found to be well-behaved under, an appropriate semiclassical approximation. To further justify the quantum equation (\ref{qh2}) in this context, we therefore need to study its behaviour under certain appropriate semiclassical limit in the standard WKB approximation. For this purpose it is much easier to handle the equation (\ref{qh1}), when it is expressed in the following form,
\begin{center}
    \be \label{qh1s} \begin{split} \bigg{[}-\frac{\hbar^2 \phi {\sqrt z}}{36\beta_{01} x}\bigg{(}\frac{\partial^2}{\partial x^2} +\frac{n}{x} \frac{\partial}{\partial x}\bigg{)}-\frac{\hbar^2}{2xz^{\frac{3}{2}}}\frac{\partial^2}{\partial \phi^2}&-{i\hbar} \frac{\partial} {\partial z}+\frac{i\hbar } {z}\bigg{(}\frac{3 \alpha_0}{\phi^2}-{\beta_{02} x^2\over 2 z^2 \phi^2}-\frac{2\gamma_0 x^2 \phi}{z^2}\bigg{)}\frac{\partial}{\partial \phi}\\&-\frac{i\hbar}{z}\bigg{(}\frac{3 \alpha_0}{\phi^3}-\frac{\beta_{02} x^2}{2 \phi^3 z^2}+\frac{\gamma_0 x^2}{z^2}\bigg{)} +\mathcal{V}(x, z, \phi)\bigg{]}\Psi=0, \end{split}\ee
\end{center}
where
\be\begin{split} \mathcal{V}(x, z, \phi)=\bigg{[}\frac{9\alpha_0^2 x}{2 \phi^4 \sqrt{z}}-\frac{3\alpha_0\beta_{02} x^3}{2 z^\frac{5}{2} \phi^4}-\frac{6\alpha_0\gamma_0 x^3}{\phi z^\frac{5}{2}}+\frac{\beta_{02}^2 x^5}{8 \phi^4 z^\frac{9}{2}}+\frac{\beta_{02}\gamma_0 x^5}{\phi z^\frac{9}{2}}+\frac{2 \gamma_{0}^2\phi^2 x^5}{z^\frac{9}{2}}+\frac{3 \alpha_{0} x}{2 \phi \sqrt{z}}+\frac{\lambda^2 \phi^2 z^\frac{3}{2}}{2 x} \bigg{]}\end{split} \ee
Equation (\ref{qh1s}) may be treated as time independent Schr{\"o}dinger equation with three variables ($x$, $z$, $\phi$), and therefore as usual, let us seek the solution of equation (\ref{qh1s}) as,

\be\label{Psi}\psi = \psi_0e^{\frac{i}{\hbar}S(x,z,\phi)},\ee
and expand $S$ in power series of $\hbar$ as,

\be\label{S} S = S_0(x,z,\phi) + \hbar S_1(x,z,\phi) + \hbar^2S_2(x,z,\phi) + .... .\ee
Now inserting the expressions (\ref{S}) and (\ref{Psi}) and their appropriate derivatives in equation (\ref{qh1s}) and equating the coefficients of different powers of $\hbar$ to zero, one obtains the following set of equations (upto second order),

\be\begin{split}
&\frac{\sqrt z \phi}{36\beta_{01} x}S_{0,x}^2 + \frac{S_{0,\phi}^2}{2xz^{\frac{3}{2}}} + S_{0,z}-\frac{1} {z}\bigg{(}\frac{3 \alpha_0}{\phi^2}-{\beta_{02} x^2\over 2 z^2 \phi^2}-\frac{2\gamma_0 x^2 \phi}{z^2}\bigg{)} S_{0,\phi} + \mathcal{V}(x,z,\phi) = 0, \label{S0}\end{split}\ee

\be\begin{split}
-\frac{i\sqrt z \phi}{36\beta_{01} x}S_{0,xx} - \frac{in\sqrt z \phi}{36\beta_{01} x^2}S_{0,x} - \frac{iS_{0,\phi\phi}}{2xz^{\frac{3}{2}}} &+ S_{1,z} + \frac{\sqrt z S_{0,x}S_{1,x}}{18\beta_{01} \phi x} + \frac{S_{0,\phi}S_{1,\phi}}{xz^{\frac{3}{2}}}-\frac{i}{z} \bigg{(}\frac{3 \alpha_0}{\phi^3}-\frac{\beta_{02} x^2}{2 \phi^3 z^2}+\frac{\gamma_0 x^2}{z^2}\bigg{)}\\&-\frac{1} {z}\bigg{(}\frac{3 \alpha_0}{\phi^2}-{\beta_{02} x^2\over 2 z^2 \phi^2}-\frac{2\gamma_0 x^2 \phi}{z^2}\bigg{)} S_{1,\phi} = 0,\label{S1}\end{split}\ee

\be\begin{split}
-i\frac{\sqrt z \phi S_{1,xx}}{36\beta_{01} x} - i\frac{n\sqrt z \phi S_{1,x}}{36\beta_{01} x^2}
+\frac{\sqrt z \phi}{36\beta_{01} x}\big(S_{1,x}^2+2 S_{0,x}S_{2,x}\big) &+ \frac{1}{2xz^{\frac{3}{2}}}\big(S_{1,\phi}^2+2S_{0,\phi}S_{2,\phi}\big)
-i\frac{S_{1,\phi\phi}}{2xz^{\frac{3}{2}}} + S_{2,z}\\&-\frac{1} {z}\bigg{(}\frac{3 \alpha_0}{\phi^2}-{\beta_{02} x^2\over 2 z^2 \phi^2}-\frac{2\gamma_0 x^2 \phi}{z^2}\bigg{)} S_{2,\phi}= 0,\label{S2}\end{split}\ee
which are to be solved successively to find $S_0(x,z,\phi),\; S_1(x,z,\phi)$ and $S_2(x,z,\phi)$ and so on. Now identifying $S_{0,x}$ as $p_x$; $S_{0,z}$ as $p_z$ and $S_{0,\phi}$ as $p_{\phi}$ one can recover the classical Hamiltonian constraint equation $H_c = 0$, given in equation (\ref{Hc}) from equation (\ref{S0}). This identifies equation (\ref{S0}) to the Hamilton-Jacobi equation. Thus, $S_{0}(x, z, \phi)$ can now be expressed as,
\be\label{S_0} S_0 = \int p_z dz + \int p_x dx + \int p_\phi d\phi, \ee
apart from a constant of integration which may be absorbed in $\psi_0$. The integrals in the above expression can be evaluated using the classical
solution for $k = 0$ presented in equation (\ref{aphi}) and equation (\ref{const}), the definition of $p_z$ and $p_{\phi}$, appearing in equation (\ref{Mom}), the definition $p_x = {Q}$, recalling the expression for ${Q}$ given in (\ref{Q}), remembering the relation, $x = \dot z$, where, $z = a^2$, and finally choosing
$n = -1$, since probability interpretation holds only for such value of $n$. Using solution (\ref{aphi}) and (\ref{const}), for $\alpha({\phi})$, $\beta_1(\phi)$, $\beta_2(\phi)$, $\gamma(\phi)$, and using the relation $x (= \dot z)$, the expressions of $p_x$, $p_z$ $p_\phi$ can be expressed in term of $x$, $z$ and $\phi$ as,
\begin{subequations}\begin{align}
&\label{allp} \hspace{5.0 cm}\alpha'=-\frac{\alpha_0}{\phi^2},\\
&\hspace{5.0 cm}x = 2{\lambda} z, \\
&\hspace{5.0 cm}p_x = \frac{36\beta_{01}\lambda x}{a_0\phi_0},\\
&\hspace{5.0 cm}p_z = -\frac{9\alpha_0\lambda z}{a_0\phi_0}-\frac{144\beta_{01}\lambda^3 z}{a_0\phi_0}+\frac{6\beta_{02}\lambda^3 z}{a_0\phi_0}+\frac{24\gamma_{0}a_{0}^2\phi_{0}^2{\lambda^3}}{\sqrt{z}},\\
&\hspace{5.0 cm}p_\phi = \frac{6\alpha_0 a_0^3\phi_0^3{\lambda}}{\phi^5}-\frac{4\beta_{02} a_0^3 \phi_0^3\lambda^3}{\phi^5}-\frac{16\gamma_0a_0^3\phi_0^3\lambda^3}{\phi^2}-\frac{a_0^3\phi_0^3\lambda}{\phi^2},
                                \end{align}\end{subequations}
Hence the integrals in (\ref{S_0}) are evaluated as,
\begin{subequations}\begin{align}
&\label{px2pphi}\hspace{5.0 cm}\int p_x dx =\frac{18\beta_{01}\lambda x^2}{a_0\phi_0}; \\
&\hspace{5.0 cm}\int p_z dz =-\frac{9\alpha_0\lambda z^2}{2a_0\phi_0}-\frac{72\beta_{01} \lambda^3 z^2}{a_0\phi_0}+\frac{3\beta_{02} \lambda^3 z^2}{a_0\phi_0}+48\gamma_{0}a_{0}^2\phi_{0}^2\lambda^3 \sqrt{z}; \\
&\hspace{5.0 cm}\int p_\phi d\phi = -\frac{3\alpha_0 a_0^3\phi_0^3{\lambda}}{2\phi^4}+ \frac{\beta_{02} a_0^3\phi_0^3{\lambda^3}}{\phi^4}+\frac{16\gamma_0a_0^3\phi_0^3\lambda^3}{\phi}+\frac{a_0^3\phi_0^3\lambda}{\phi}.
\end{align}\end{subequations}
\noindent
Therefore, explicit form of $S_0$ in terms of $z$ is found as,

\be\label{S0f} S_0 = \frac{2\alpha_0\lambda z^2}{a_0\phi_0}+16\gamma_{0}a_{0}^2\phi_{0}^2\lambda^3\sqrt{z}.\ee
For consistency, one can trivially check that the expression for $S_0$ (\ref{S0f}) so obtained, satisfies equation (\ref{S0}) identically. In fact it should, because, equation (\ref{S0}) coincides with Hamiltonian constraint equation (\ref{Hc}) for $k = 0$, and therefore is the Hamilton-Jacobi equation, as already mentioned. Moreover, one can also compute the zeroth order on-shell action (\ref{Sc3}). For example, using the relation of $ \alpha_0,~
 \beta_{01}, \beta_{02}$ and $\gamma_0$ and  classical solution (\ref{aphi}) one may express all the variables in terms of $t$ and substitute in the action (\ref{Sc3}) to obtain,

\be\label{Acl}A=A_{cl}=\int\left[\frac{8\alpha_0 a_0^3\lambda^2}{\phi_0}e^{{4\lambda}t}+16\gamma_0a_0^3\phi_0^2{\lambda}^4e^{\lambda t}\right]dt.\ee
Integrating we have, \be\label{4.87} A=A_{cl}=\frac{2\alpha_0a_0^3\lambda}{\phi_0}e^{{4\lambda}t}+16\gamma_0a_0^3\phi_0^2{\lambda}^3e^{\lambda t},\ee
which is the same as we obtained in (\ref{S0f}), which is a consistency check. At this end, the wave function therefore reads as,

\be\label{psif} \Psi = \Psi_0 e^{\frac{i}{\hbar}\left[ \frac{2\alpha_0\lambda z^2}{a_0\phi_0}+16\gamma_{0}a_{0}^2\phi_{0}^2\lambda^3\sqrt{z}\right]}.\ee

\subsection{\bf{First order approximation}}
Now for $n=-1$, equation (\ref{S1}) may be expressed as,
\be\begin{split}
-\frac{\sqrt z\phi}{36\beta_{01} x}\Big({}iS_{0,xx}-2S_{0,x}S_{1,x} - \frac{i}{ x}S_{0,x}\Big{)} &-\frac{1}{2xz^{\frac{3}{2}}} \Big{(}i{S_{0,\phi\phi}}-2S_{0,\phi}S_{1,\phi}\Big{)}-\frac{1} {z}\bigg{(}\frac{3 \alpha_0}{\phi^2}-{\beta_{02} x^2\over 2 z^2 \phi^2}-\frac{2\gamma_0 x^2 \phi}{z^2}\bigg{)} S_{1,\phi}\\& -\frac{i}{z} \bigg{(}\frac{3 \alpha_0}{\phi^3}-\frac{\beta_{02} x^2}{2 \phi^3 z^2}+\frac{\gamma_0 x^2}{z^2}\bigg{)}+ S_{1,z} = 0, \label{S11}\end{split}\ee
Using the expression of $S_0$ from (\ref{S0f}), we can find $S_{1,z}$ from the above equation as,
\be\label{S1z} S_{1,z} = i \left(\frac{{C_1\sqrt z} +\frac{C_2}{z}+\frac{C_3}{z^\frac{5}{2}}}{D_1z^\frac{3}{2}+\frac{D_2}{z^\frac{3}{2}}+D_3}\right),\ee
where, $C_1=\frac{9\alpha_0}{ a_0^3\phi_0^3},~~C_2=12\gamma_0 \lambda^2,~~ C_3=-\frac{\gamma_0 a_0^3 \phi_0^3}{24 \beta_{01}},~~D_1={6\alpha_0\over a_0^3 \phi_0^3},~~D_2=\frac{\gamma_0 a_0^3 \phi_0^3}{18\beta_{01}}$, and $D_3=\left(1+\frac{\alpha_0}{36\beta_{01}\lambda^2}\right)$ are all constants. The above equation \eqref{S1z} may be integrated in principle, and  $S_1$ may be expressed in the form,

\begin{center}
    \be S_1=iF(z). \ee
\end{center}
Therefore the wave function up to first-order approximation reads as,
\begin{center}
    \be \label{Psi} \Psi = \psi_{01} e^{\frac{i}{\hbar}\left[\frac{2\alpha_0\lambda z^2}{a_0\phi_0}+16\gamma_{0}a_{0}^2\phi_{0}^2\lambda^3\sqrt{z}\right]},\ee
\end{center}
where, \be \psi_{01}=\psi_0e^{-F(z)},\ee
which only tells upon the pre-factor keeping the exponent part unaltered. We have therefore exhibited a technique to find the semiclassical wavefunction, on-shell. One can proceed further to find higher order approximations. Nevertheless, it is clear that higher order approximations too, in no way would affect the form of the semiclassical wavefunction, which has been found to be oscillatory around the classical inflationary solution. Let us mention that Hartle proposed a criterion for the selection of classical trajectories \cite{HC}. The proposal is to look for the peaks of the wave function of the universe. It states: if it is strongly peaked, there exists correlations among the geometrical and matter degrees of freedom, and the emergence of classical trajectories (i.e. the universe) is expected. However, if it is not peaked, correlations are lost. The above statement works as well for general classes of `Extended Theories of Gravity' and is found to be conformally  preserved \cite{Cap1}. Further, it has also been revealed that different minisuperspace cosmological models, such as non-minimally coupled scalar tensor theory, $F(R)$ theory, and even theories higher than fourth order, show up oscillatory behaviour of the wave function \cite{Cap2}. Likewise, since the semiclassical wavefunction \eqref{Psi} exhibits oscillatory behaviour and is strongly peaked around classical de-Sitter solution \eqref{aphi}, Hartle criterion is fulfilled. This proves that the quantum counterpart of the action \eqref{Scf} produces a reasonably viable theory.

\section{Inflation under Slow Roll Approximation:}

Since its advent, the theory of inflation \cite{Guth, Linde1, Linde2, Staro, Stein} has been established as a scenario rather than a model. Inflation is a quantum phenomena and it must have occurred just around Planck's era. In the previous sub-section we have mentioned that if a quantum theory admits a viable semiclassical approximation, then most of the important physics may be extracted from the classical action itself. Having proved the viability of the action \eqref{A} in the quantum domain, we have observed that the semiclassical wavefunction oscillates around a de-Sitter solution. This indicates that the universe, according to the present model, enters an inflationary regime, soon after Planck's era. We therefore now proceed to test inflation with currently released data sets from Planck \cite{pd1, pd2}, in view of the classical field equations. Although, it appears that starting from an action containing a generic function, $F(R, \mathcal{G})$, all the curvature budget of effective gravitational theories (other than derivatives of curvature variants) are exhausted \cite{Cap3}, nevertheless an action containing ($\alpha R + \beta G^2$) term show up ceertain pathologies, some of which may be resolved in the presence of cosmological constant term \cite{We13}. Here, we show that the presence of higher order curvature invariant terms in the presence of Gauss-Bonnet-dilatonic coupling indeed lead to appreciably nice result in the context of inflation. For a complicated theory such as the present one, it is of-course a very difficult job. However, we follow a unique technique to make things look rather simple. Let us first rearrange the ($^0_0$) and the $\phi$ variation equations of Einstein, viz., (\ref{00}) and (\ref{phivariation}) respectively as,

\be\label{00H}\begin{split} - 6\alpha \mathrm{H}^2 - 6\alpha'{\dot\phi}\mathrm{H} - 36\beta_1 \mathrm{H}^4 &\bigg{[}4\bigg{(}1+\frac{\dot {\mathrm{H}}}{\mathrm{H}^2}\bigg{)}+4\frac{\dot{ \mathrm{H}}}{\mathrm{H}^2}\bigg{(}1+\frac{\dot{\mathrm{H}}}{\mathrm{H}^2}\bigg{)} +2\bigg{(}\frac{\ddot {\mathrm{H}}}{\mathrm{H}^3}-2\frac{{\dot {\mathrm{H}}}^2}{\mathrm{H}^4}\bigg{)}-\bigg{(}1+\frac{\dot {\mathrm{H}}} {\mathrm{H}^2} \bigg{)}^2-3\bigg{]}\\& - 72\beta'_1{\dot\phi} \mathrm{H}^3\bigg{[}\bigg{(}1+\frac{\dot {\mathrm{H}}}{\mathrm{H} ^2}\bigg{)}+ 1\bigg{]}+12\beta'_2{\dot\phi} \mathrm{H}^3 - 24\gamma'\dot\phi \mathrm{H}^3 + \frac{\dot\phi^2}{2}+V=0, \end{split} \ee

\be \label{phivar} \begin{split}&\ddot\phi +3\mathrm{H}\dot\phi + V' =\\& 6\alpha'\mathrm{H}^2\bigg{[}\bigg{(}1+\frac{\dot {\mathrm{H}}} {\mathrm{H}^2} \bigg{)}+1\bigg{]}+36\beta'_1 \mathrm{H}^4\bigg{[} \bigg{(}1+\frac{\dot{\mathrm{H}}}{\mathrm{H}^2}\bigg{)}^2+2\bigg{(}1+\frac{\dot {\mathrm{H}}}{\mathrm{H}^2}\bigg{)}+1\bigg{]}-12\beta'_2\mathrm{H}^4+24\gamma'\mathrm{H}^4\bigg{(}1+\frac{\dot {\mathrm{H}}}{\mathrm{H}^2}\bigg{)},
\end{split}\ee
where, $\mathrm{H}={\dot a\over a}$ denotes the expansion rate, which is assumed to be slowly varying, and replaces the constant $\lambda$, appeaing in the exponent of the inflationary solutions in \eqref{aphi} of the classical field equations (\ref{00}) and (\ref{phivariation}) obtained in standard de-Sitter form. Now, instead of standard slow roll parameters, we introduce a hierarchy of Hubble flow parameters \cite{CH1, CH2, CH3, CH4, We1, We2} in the following manner, which appears to be much suitable to handle higher order theories. Firstly, the background evolution of the theory under consideration is described by a set of horizon flow functions (the behaviour of Hubble distance during inflation) starting from,

\be \label{dh}\epsilon_0=\frac{d_{\mathrm{H}}}{d_{\mathrm{H}_i}},~ \text{where}~~ d_{\mathrm{H}}=\mathrm{H}^{-1},
\ee
where, $d_{\mathrm{H}} = \mathrm{H}^{-1}$ is the Hubble distance, also called the horizon in our chosen units. We use suffix `$i$' to denote the era at which inflation was initiated. Now hierarchy of functions is defined in a systematic way as,

\begin{center}
    \be \label{el} \epsilon_{l+1}=\frac{d\ln|\epsilon_l|}{d \mathrm{N}},~~l\geq 0. \ee
\end{center}
In view of the definition $\mathrm{N}=\ln{\frac{a}{a_i}}$, implying $\mathrm{\dot N}=\mathrm{H}$, one can compute $\epsilon_1=\frac{d\ln{d_{\mathrm{H}}}}{d\mathrm{N}},$ which is the logarithmic change of Hubble distance per e-fold expansion $\mathrm{N}$, and is the first slow-roll parameter $\epsilon_1=\dot{d_{\mathrm{H}}}=-\frac{\dot {\mathrm{H} }}{\mathrm{H}^2}$. This signals that the Hubble parameter almost remains constant during inflation. The above hierarchy allows one to compute $\epsilon_2=\frac{d\ln{\epsilon_1}}{d\mathrm{N}}=\frac{1}{\mathrm{H}}\frac{\dot\epsilon_1}{\epsilon_1},$ which implies $\epsilon_1\epsilon_2=d_{\mathrm{H}} \ddot{d_{\mathrm{H}}} =-\frac{1}{\mathrm{H}^2}\left(\frac{\ddot {\mathrm{H}}}{\mathrm{H}}-2\frac{\dot {\mathrm{H}}^2}{\mathrm{H}^2}\right)$. In the same manner higher slow-roll parameters may be computed. Equation (\ref{el}) essentially defines a flow in space with cosmic time being the evolution parameter, which is described by the equation of motion

\be\label{el1}\epsilon_0\dot\epsilon_l-\frac{1}{d_{\mathrm{H}_i}}\epsilon_l\epsilon_{l+1}=0,~~~~l\geq 0.\ee
In view of the slow-roll parameters, equations (\ref{00H}) and (\ref{phivar}) may therefore be expressed as,

\be \label{hir1}\begin{split} -6\alpha \mathrm{H}^2-6\alpha'\dot\phi\mathrm{H}-36\beta \mathrm{H}^4\left[3\big(1-\epsilon_1\big)^2-2\big(1+\epsilon_1 \epsilon_2 \big)-1\right]&-72\beta'_1\dot\phi\mathrm{H} ^3\left[\big(1- \epsilon_1\big)+1\right]\\&+12\beta'_2{\dot\phi} \mathrm{H}^3-24\gamma'\dot\phi \mathrm{H}^3+\Big(\frac{\dot\phi^2}{2}+V\Big)=0, \end{split}\ee
and
\be \label{phi2} \ddot\phi +3\mathrm{H}\dot\phi=-V'+6\alpha'\mathrm{H}^2\left[3-\big(1+\epsilon_1\big)\right]
+{36\beta'_1\mathrm{H}^4}\left[\big(1-\epsilon_1\big)^2+2\big(1-\epsilon_1\big)+1\right]-12\beta'_2 \mathrm{H}^4 +24\gamma' \mathrm{H}^4\big(1- \epsilon_1\big), \ee
respectively, which may therefore be approximated using the slow roll hierarchy to,

\be \label{hir11}\begin{split} 6\alpha \mathrm{H}^2 = \frac{\dot\phi^2}{2} + \left[V - \big(6\alpha'\dot\phi\mathrm{H} + 144\beta'_1\dot\phi\mathrm{H} ^3 -12\beta'_2{\dot\phi} \mathrm{H}^3 + 24\gamma'\dot\phi \mathrm{H}^3\big)\right], \end{split}\ee
and
\be \label{phi22} \ddot\phi +3\mathrm{H}\dot\phi + \left[V' - \big(12\alpha'\mathrm{H}^2+{144\beta'_1\mathrm{H}^4}-12\beta'_2 \mathrm{H}^4 +24\gamma' \mathrm{H}^4\big)\right] = 0. \ee
Before imposing the standard slow roll conditions, viz. $|\ddot \phi| \ll 3\mathrm{H}|\dot \phi|$ and $\dot\phi^2 \ll V(\phi)$, we try to reduce equations \eqref{hir11} and \eqref{phi22} in a much simpler form. For example redefining an effective potential ($U$) as,

\be\label{U1} U = V -12\mathrm{H}^2(\alpha+12\mathrm{H}^2\beta_1-\mathrm{H}^2\beta_2+2\mathrm{H}^2\gamma),\ee
equation (\ref{phi22}) takes the standard form of Klein-Gordon Equation,

\be \label{KG}\begin{split} \ddot\phi +3\mathrm{H}\dot\phi + U'=0. \end{split} \ee
In view of the reduced equation \eqref{KG}, it is now quite apparent that the evolution of the scalar field is driven by the re-defined potential gradient $U' = {dU\over d\phi}$, subject to damping by the Hubble expansion $3\mathrm{H} \dot \phi$, as in the case of single field equation, while the potential $U(\phi)$ carries all the information in connection with the coupling parameters of generalised higher order action under consideration. Further assuming,

\be\label{U2} U = V - 6\mathrm{H}\dot \phi\left(\alpha' + 24 \mathrm{H}^2 \beta'_1-2\mathrm{H}^2 \beta'_2+ 4 \mathrm{H}^2 \gamma'\right),\ee
equation \eqref{hir11} may be reduced to the following simplified form, viz,

\be\label{Fried} 6\alpha\mathrm{H}^2 = \frac{\dot\phi^2}{2} + U(\phi),\ee
which is simply the Friedmann equation with non-minimal coupling $\alpha$. It is important to mention that, the two choices on the redefined potential $U(\phi)$ made in \eqref{U1} and \eqref{U2}, do not confront in any case and may be proved to be consistent as demonstrated underneath. During slow roll, the Hubble parameter $\mathrm{H}$ almost remains unaltered. Thus replacing $\mathrm{H}$ by $\lambda$, and using the forms of the parameters $\alpha(\phi)$ presented in \eqref{aphi}, along with $\beta_1(\phi)$, $\beta_2(\phi)$ and $\gamma(\phi)$ assumed in \eqref{const}, the two relations \eqref{U1} and \eqref{U2} lead to the following first order differential equation on $\phi$,

\be \label{Consistency} \left(\frac{\phi^3 - 6\alpha_0}{\phi^4}\right) d\phi = \frac{\lambda}{2} dt,\ee
which can immediately be integrated to yield,

\be \label{phidecay}\ln{\phi} + \frac{2\alpha_0}{\phi^3} = \frac{\lambda}{2}(t - t_0).\ee
Clearly, if $\phi$ is not too large, $\ln{\phi}$ remains subdominant, and $\phi$ falls-of with time, as expected during inflationary regime. Thus for the classical de-Sitter solution, the forms of the coupling parameters given in \eqref{aphi} and \eqref{const}, also leads to the above solution of $\phi$, instead of the exponentially fall-of one, presented in \eqref{aphi}. Having proven consistency of our assumptions, which reduce the field equations to the non-minimally coupled Friedmann equation with a single scalar field, we can now enforce the standard slow-roll conditions $\dot\phi^2\ll U$, and $|\ddot\phi|\ll 3\mathrm{H}|\dot\phi|$, on equations (\ref{Fried}) and (\ref{KG}), which thus finally reduce to,

\be\label{H2}{6\alpha}\mathrm{H}^2\simeq U, \ee
and
\be\label{Hphi} 3\mathrm{H}\dot\phi \simeq - U'. \ee
Now, combining equations (\ref{H2}) and (\ref{Hphi}), it is possible to show that the potential slow roll parameter $\epsilon$ equals the Hubble slow roll ($\epsilon_1$) parameter under the condition,
\be\label{SR} \epsilon = - {\dot {\mathrm{H}}\over \mathrm{H}^2} = \alpha\left({U'\over U}\right)^2 - \alpha'\left({U'\over U}\right);\hspace{0.4 cm} \eta = 2 \alpha \left({U''\over U}\right),\ee
while $\eta$ remains unaltered. Further, since $\frac{\mathrm{H}}{\dot\phi}=-{U\over2\alpha U'}$, therefore, the number of e-folds, at which the present Hubble scale equals the Hubble scale during inflation, may be computed as usual in view of the following relation:

\be\label{Nphi} \mathrm{N}(\phi)\simeq \int_{t_i}^{t_f}\mathrm{H}dt=\int_{\phi_i}^{\phi_f}\frac{\mathrm{H}}{\dot\phi}d\phi\simeq \int_{\phi_f}^{\phi_i}\Big{(}\frac{U}{2\alpha U'}\Big{)}d\phi,\ee
where, $\phi_i$ and $\phi_f$ denote the values of the scalar field at the beginning $(t_i)$ and the end $(t_f)$ of inflation. Thus, slow roll parameters reflect all the interactions, as exhibited earlier \cite{GB1, GB2, GB3}, but here only via the redefined potential $U(\phi)$.\\

Let us now consider the potential in its most standard form, viz, $V(\phi) = {1\over 2}\lambda^2\phi^2$, and also the forms of $\alpha(\phi)~\mathrm{in}~ $\eqref{aphi}, $\beta_1(\phi)$, $\beta_2(\phi)$ and $\gamma(\phi)$ in \eqref{const}, which satisfy classical de-Sitter solutions, in order to compute inflationary parameters numerically. So at first, it is necessary to find the form of the re-defined potential $U(\phi)$. As mentioned, during inflation the Hubble parameter remains almost constant, and therefore while computing $U(\phi)$, one can replace it by the constant $\lambda$, without any loss of generality. Thus,

\be 12\mathrm{H}^2\left(\alpha+12\mathrm{H}^2\beta_1-\mathrm{H}^2\beta_2+2\mathrm{H}^2\gamma\right) \approx -\frac{\mathrm{H}^2\phi^2}{2},~~\mathrm{such~ that},~~ U = {1\over 2} {m^2\phi^2},~~\mathrm{where},~~ m^2 = \lambda^2+\mathrm{H}^2 \approx 2\lambda^2.\ee
In view of the above quadratic form of the re-defined potential, the slow roll parameters $\epsilon$ and $\eta$ \eqref{SR} and the number of e-folding $\mathrm{N}$ \eqref{Nphi} take the following forms,

\be\label{epseta}\epsilon = \frac{6\alpha_0}{\phi^3},\hspace{0.5 cm}\eta = \frac{4\alpha_0}{\phi^3}, \hspace{0.5 cm}\mathrm{N} = {1\over 4 \alpha_0}\int_{\phi_f}^{\phi_i} \phi^2 d\phi = {1\over 12\alpha_0}(\phi_i^3 - \phi_f^3).\ee
Further, comparing expression for the primordial curvature perturbation on super-Hubble scales produced by single-field inflation, $P_\zeta(k)$ with the primordial gravitational wave power spectrum $P_t(k)$, one obtains the tensor-to-scalar ratio for single-field slow-roll inflation, $r = {P_t(k)\over P_\zeta(k)} = 16\epsilon$, while, the scalar tilt, conventionally defined as $n_s-1$, may be expressed by the spectral index as, $n_s = 1 - 6\epsilon + 2\eta$. In view of all these expressions we compute the inflationary parameters and present them for different values of the parameter $\alpha_0$  in table 1. We also present respective $n_s$ versus $r$ plots in figure 1.\\

\begin{figure}

   \begin{minipage}[h]{0.47\textwidth}
      \centering
      \begin{tabular}{|c|c|c|c|c|}
     \hline\hline
      $\alpha_0$ in $M_P^3$ & $\phi_f$ in $M_P$ & $n_s$ & $r$ & $\mathrm{N}$\\
      \hline
       0.066 & 0.734 & 0.9767 & 0.0796 & 99\\
      0.068 & 0.741 & 0.9760 & 0.0821 & 96\\
      0.070& 0.749 & 0.9753 & 0.0845& 94\\
      0.072 & 0.756 & 0.9746 & 0.0869 &91\\
      0.074 & 0.763 & 0.9739 & 0.0893 & 89\\
      0.076& 0.769 & 0.9732& 0.0917 & 86\\
      0.078& 0.776 & 0.9725& 0.0941 & 84\\
      0.080& 0.783 & 0.9718 & 0.0965 & 82\\
      0.082& 0.789 & 0.9711 & 0.0990 & 80\\
      0.084& 0.795 & 0.9704 & 0.1014 & 78\\
      \hline\hline
    \end{tabular}
      \captionof{table}{Data set for the inflationary parameters taking $\phi_i=4.3 M_P$ while $\alpha_0$ is varied.}
      \label{tab:table1}
   \end{minipage}%
   \hfill%
   \begin{minipage}[h]{0.47\textwidth}
      \centering
      \includegraphics[width=0.9\textwidth]{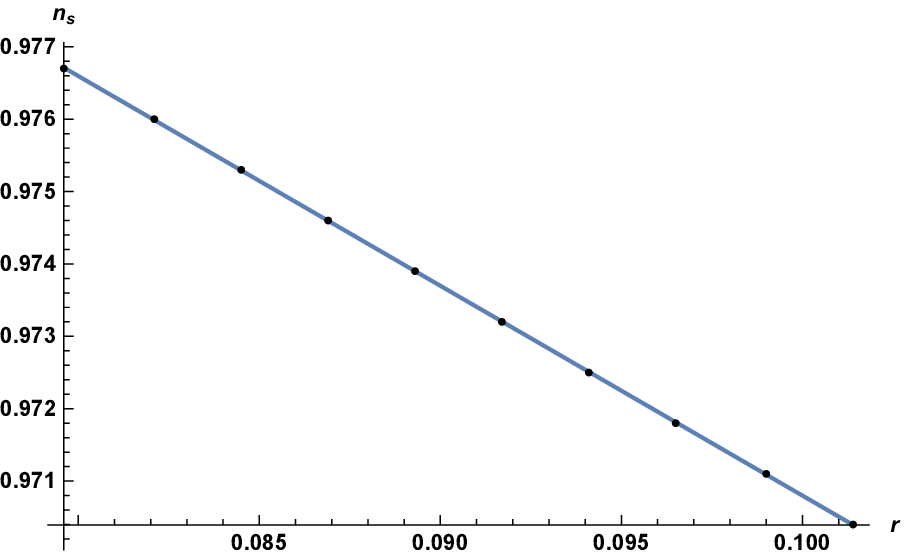}
      \caption{This plot depicts the variation of $n_s$ with $r$, varying $\alpha_0$.}
      \label{fig:fig1}
   \end{minipage}%

\end{figure}

\noindent
Table 1 depicts that under the variation of $\alpha_0$ within the range $0.066 M_P^3\le\alpha_0 \le 0.084M_P^3$, the spectral index of scalar perturbation lies within the range $0.970 \le n_s \le 0.977$, which shows excellent agreement, while the scalar to tensor ratio lies within the range $0.079\le r \le 0.101$, showing reasonably good agreement with the recently released data \cite{pd1, pd2}. Further, since $m^2$ remains arbitrary, the energy scale of inflation maybe chosen at sub-Planckian domain, e.g., $\mathrm{H}_* \approx 10^{-5} M_P$, in conformity with single scalar field inflation. This also validates the semiclassical approximation. However, the number of e-folds varies within the range $78 \le \mathrm{N} \le 99$, which is although sufficient to solve the horizon and flatness problems, is considerably large than usual. Attempt to reduce it tells upon the scalar to tensor ratio, since it increases, as is clearly visible from the table 1. For the sake of visualization we present the spectral index of scalar perturbation versus the scalar to tensor ratio plot in figure 1.\\

\noindent
Although the theory under consideration is highly complicated, we have been able to reduce the system of field equations considerably to study inflation, without using additional flow parameters. In fact it is also possible to demonstrate that it does not also suffer from graceful exit problem. The scalar decays as $\phi \sim t^{-3}$ \eqref{phidecay}, and quickly falls below the Planck's mass, $\phi < M_p$. To exhibit the fact that as $\phi \ll M_p$ the field approaches an oscillatory solution, let us express equation \eqref{Fried} as

\be 3\mathrm{H}^2 = {1\over 2\alpha}\left({1\over 2}\dot\phi^2 + {1\over 2} m^2 \phi^2\right),\ee
where, $U = {1\over 2} m^2 \phi^2$, and $m^2 \approx 2 \lambda^2$. In view of the expression of $\alpha(\phi) = {\alpha_0\over \phi}$, the above equation reads as,

\be \label{Hubble} {3\mathrm{H}^2\over m^2} = {3\mathrm{H}^2\over 2\lambda^2} = {\phi\over 4\alpha_0}\left({\dot\phi^2\over 2\lambda^2} + \phi^2\right).\ee
Note that for single scalar field, the above equation reads as $3\mathrm{H}^2 = {1\over 2M_p^2}(\dot\phi^2 + m^2 \phi^2)$. At the end of inflation, ${\phi\over 4\alpha_0} \sim {2\over M_p^2}$, according to the present data set. Once the Hubble rate falls below $\sqrt {2}\lambda$, the left hand side may be neglected, and the equation \eqref{Hubble} may be approximated to,

\be \dot\phi^2 \approx -2\lambda^2 \phi^2,\ee
which exhibits oscillatory behaviour of $\phi \sim e^{i {\sqrt{2}\lambda}~ t}$. The field therefore starts oscillating many times over a Hubble time, driving a matter-dominated era at the end of inflation.

\noindent
The number of e-folds as exhibited in table 1, being higher than usual, suggests that the model might require some non-standard post-inflationary cosmology. This is related with the stabilization issue associated with the dilaton, since generic string models suffer from dilaton runaway problem \cite{DR}. However, this pathology might be controlled considering an additive constant in the potential, which does not affect the solutions to the classical field equations, and can be absorbed in the potential function itself. However, inflationary parameters $\epsilon,~\eta$ and the number of e-folds $\mathrm{N}$, depend on the potential function itself and are likely to modify the situation. For example, if we take $U = {1\over 2}m^2 \phi^2 - u_0$, where, $u_0$ is a constant, the inflationary parameters read as,

\be\label{epseta2}\epsilon = \frac{4m^4\alpha_0\phi}{(m^2\phi^2-2u_0)^2}+\frac{2m^2\alpha_0}{(m^2\phi^3-2 u_0 \phi )},\hspace{0.5 cm}\eta = \frac{4m^2\alpha_0}{m^2\phi^3-2 u_0 \phi  }.\ee
\be\label{N}\mathrm{N} = {1\over 4\alpha_0}\int_{\phi_f}^{\phi_i}{(m^2\phi^2-2 u_0 )\over m^2}d\phi = {1\over 12\alpha_0}(\phi_i^3 - \phi_f^3)-{u_0\over 2m^2\alpha_0}(\phi_i-\phi_f).\ee

\noindent
In table 2, we have fixed $u_0 = 1.0 {M^4_P}$ and varied $\alpha_0$ within the range ($0.80\times10^{-5})M_P^3\le\alpha_0 \le (1.25\times10^{-5}) M^3_P$. One can observe that the spectral index of scalar perturbation and the scalar to tensor ratio lie within the range $0.967 \le n_s \le 0.979$ and $0.057\le r \le 0.089$ respectively, while the number of e-folds varies within the range $46 \le \mathrm{N} \le 72$, which are in excellent agreement with the recently released data \cite{pd1, pd2}. In table 3 on the contrary, we fix $\alpha_0 = (1.25\times10^{-5}){M^3_P}$, and allow variation of $u_0$ within the range $0.986 M^4_P \le u_0 \le 1.000 M^4_P $. As a result, the spectral index of scalar perturbation and the scalar to tensor ratio lie within the range $0.967 \le n_s \le 0.979$ and $0.056\le r \le 0.089$ respectively, while the number of e-folds varies within the range $46 \le \mathrm{N} \le 74$. Here again for the sake of visualization we present the spectral index of scalar perturbation versus the scalar to tensor ratio, associated with table 2 and table 3 in the plots figure 2 and figure 3. In the process, the problem with large number of e-folds, which could make the universe sufficiently cold, is circumvented. It is important to mention the fact that: since the computed sum of zero point energies of all fields, viz. the vacuum energy density,

\be <T_{00}>_{\mathrm{vac}} = \rho_\mathrm{vac} \approx 10^{73} GeV^4,\ee
while the additive constant here is also, $u_0 \sim 1 M_P^4 \approx 10^{73} GeV^4$, so $u_0$ is essentially the vacuum energy density. This suggests that, the problem is circumvented by the addition of a cosmological constant in disguise. It can also be shown as before, that the model admits graceful exit from inflation.

\begin{figure}

   \begin{minipage}[h]{0.47\textwidth}
      \centering
      \begin{tabular}{|c|c|c|c|c|}
      \hline\hline
      $\alpha_0$ in ${10^{-5} M^3_P}$ & $\phi_f$ in $M_P$ & $n_s$ & $r$ & $\mathrm{N}$\\
      \hline
      1.25& 1.48782 & 0.9671 & 0.08911 & 46\\
      1.20& 1.48788 & 0.9685 & 0.08554 & 48\\
      1.15& 1.48794 & 0.9698 & 0.08198 & 50\\
      1.10& 1.48800 & 0.9712 & 0.07841 & 53\\
      1.05& 1.48806 & 0.9723 & 0.07485 & 55\\
      1.00& 1.48812 & 0.9737 & 0.07129 & 58\\
      0.95& 1.48819 & 0.9750 & 0.06772 & 60\\
      0.90& 1.48826 & 0.9763 & 0.06416 & 64 \\
      0.85& 1.48832 & 0.9777 & 0.06059 & 68 \\
      0.80& 1.48840 & 0.9790 & 0.05703 & 72 \\
       \hline\hline
    \end{tabular}
      \captionof{table}{Data set for the inflationary parameters taking $\phi_i=1.53 M_P$;~$m^2 = 0.9 {M^2_P}$;~$u_0=1.0 {M^4_P}$ and ~ varying $\alpha_0$.}
      \label{tab:table2}
   \end{minipage}%
   \hfill%
   \begin{minipage}[h]{0.47\textwidth}
      \centering
      \includegraphics[width=0.9\textwidth]{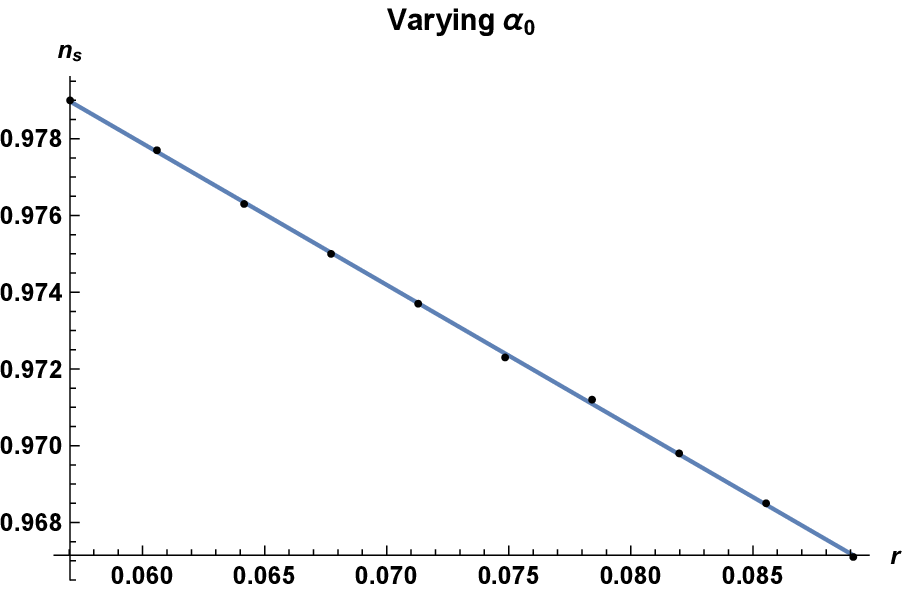}
      \caption{This plot depicts the variation of $n_s$ with $r$, varying $\alpha_0$.}
      \label{fig:fig2}
   \end{minipage}%

\end{figure}
\begin{figure}

   \begin{minipage}[h]{0.47\textwidth}
      \centering
      \begin{tabular}{|c|c|c|c|c|}
     \hline\hline
      $u_0$~ in $M^4_P$ & $\phi_f$ in $M_P$ & $n_s$ & $r$ & $\mathrm{N}$\\
      \hline
      1.000 & 1.48782 & 0.9671 & 0.08911 & 46 \\
      0.998 & 1.48633 & 0.9695 & 0.08287 & 50 \\
      0.996 & 1.48483 & 0.9715 & 0.07726 & 53\\
      0.994 & 1.48334 & 0.9734 & 0.07222 & 57\\
      0.992 & 1.48184 & 0.9751 & 0.06765 & 61\\
      0.990 & 1.48034 & 0.9766 & 0.06351 & 65\\
      0.988 & 1.47884 & 0.9780 & 0.05974 & 70\\
      0.986 & 1.47734 & 0.9793 & 0.05630 & 74\\

      \hline\hline
    \end{tabular}
      \captionof{table}{Data set for the inflationary parameters taking $\phi_i=1.53 M_P$;~$m^2 = 0.9 {M^2_P}$;~$\alpha_0=1.25\times10^{-5} {M^3_P}$~  and  ~ varying $u_0$.}
      \label{tab:table3}
   \end{minipage}%
   \hfill%
   \begin{minipage}[h]{0.47\textwidth}
      \centering
      \includegraphics[width=0.9\textwidth]{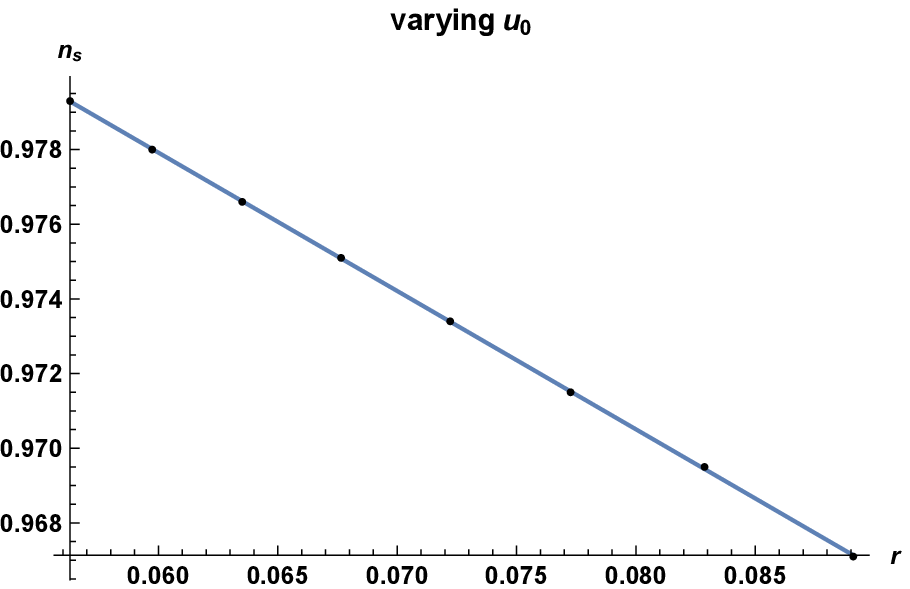}
      \caption{This plot depicts the variation of $n_s$ with $r$, varying $u_0$.}
      \label{fig:fig3}
   \end{minipage}%
\end{figure}

\section{\bf{Concluding remarks}}

We first make it clear that, neither do we have any obligation to any particular model, nor any intention to promote a model. Our aim is simply to test the viability of different models in the cosmological context, particularly, in the context of the evolution of the early universe, as in the present manuscript. For this purpose, perhaps the most general action upto curvature squared term (apart from a pure cosmological constant term), in the minisuperspace model guided by cosmological principle, has been considered to open up a possible window through which a glimpse of the very early universe might enable us to acquire some intuitive picture with certain insights. The fact that the action admits vacuum de-Sitter solution assures the viability of the action as a first check. Interestingly, the solutions so obtained admit identical forms as in \cite{We11} without $R_{\mu\nu}^2$ and Gauss-Bonnet-dilatonic coupled terms. It may be noticed that the de-Sitter solution restricts the potential to the standard quadratic form for which $\big|{V'(\phi)\over V(\phi)}\big| \ll 0$ for a wide range of the value of the scalar field $\phi$. Therefore, one does not require additional assumption, such as $\dot \phi \approx 0$, as in the case of minimally coupled scalar-tensor theory of gravity, which is procured only under appropriate initial/boundary condition on the wavefunction. Phase-space structure of the action has been presented executing Modified Horowitz' Formalism, and the Hamiltonian is quantized following the standard canonical quantization scheme. The effective Hamiltonian operator turns out to be  hermitian, which is the necessary and sufficient condition for unitarity. Thus, despite the presence of $R_{\mu\nu}^2$ term, the action is non-perturbatively well behaved, while the operator ordering index $n$ is fixed once and forever to $n = -1$, to render hermiticity of the effective Hamiltonian. Although as usual, time ceases to exist, nevertheless, an internal parameter (the proper volume) plays its role, which allows to establish the standard probabilistic interpretation. Based on the only primitive prediction that the universe is approximately classical when it is large, the semiclassical approximation (on-shell) has been performed. The approximate wave-function so obtained admits Hartle' criterion, since it exhibits oscillatory behaviour about classical inflationary solution, i.e. it is peaked about the de-Sitter solution to the classical Einstein equations. This indicates that the quantum equation is classically admissible. This is essentially the first part of the present work, which reveals the fact that the present model enters into an inflationary regime soon after Planck's era. It therefore necessitates the study of inflation, which is performed in section 4.\\

In the presence of dilatonic coupling, a combined hierarchy of Hubble and Gauss–Bonnet flow parameters is usually required to introduce, since additional condition apart from the standard slow roll, is necessary \cite{We8, We9, CH1, CH2, CH3, CH4}. Even, in the case of a non-minimally coupled scalar-tensor theory of gravity in the presence of scalar curvature squared term with constant coupling (without Gauss-Bonnet term), again a combined hierarchy of Hubble and non-minimal flow parameters is required \cite{We11}. It therefore appears that additional conditions are necessary corresponding to the number of coupling parameters present in the theory. The novelty of the present study is that: only a combined hierarchy of Hubble flow parameter has been found to be enough to evaluate inflationary parameters, and even the hierarchy of the Gauss–Bonnet flow parameter is not required at all. In fact, through the redefinition of the effective potential, the classical field equations reduce to the Friedmann equation with non-minimal coupling and the standard Klein-Gordon equation. \\

Although the inflationary parameters (spectral index and scalar to tensor ratio) are in good agrement with the latest released data set from Planck (Table-1), the number of e-folds is too large, which cools the universe below normal. The model therefore appears to suffer from the reheating issue. An additive constant in the potential removes the pathology, and the parameters are found to be in excellent agreement with the available data set \cite{pd1, pd2}, as depicted in figure-2 and figure-3. The additive constant $u_0 \sim 1 M_P^4$ suggests that, it is essentially the vacuum energy density, since $\rho_{\mathrm{vac}} \sim 10^{73} GeV^4$, and so the inflation is essentially driven by a cosmological constant. The model admits graceful exit from inflation.

\appendix

\section{Canonical formulation by Dirac's constraint analysis:}

The aim of the appendix is to show that if one initiates Dirac formalism of constraint analysis only after taking care of the divergent terms appearing in the action, then it leads to identical Hamiltonian (\ref{Hc}) as obtained following `Modified Horowitz' Formalism'. We therefore integrate the appropriate terms appearing in the action (\ref{Sc1}) by parts, to express the point Lagrangian in view of (\ref{Sc2}), in the following form,

\begin{center}\be \label{A1}\begin{split} L = &\bigg{[}{\bigg{(}-\frac{3\alpha'\dot\phi\dot z\sqrt z}{N}-\frac{3\alpha\dot z^2}{2N\sqrt z}+6kN\alpha\sqrt z\bigg{)}}+\frac{9\beta_1}{\sqrt z}\bigg{(}\frac{\ddot z^2}{N^3}-\frac{2\dot z \ddot z \dot N}{N^4}+\frac{\dot z^2\dot N^2}{N^5}+\frac{2k{\dot z}^2}{Nz}+4k^2N \bigg{)}\\&-\frac{36\beta_1' k \dot z \dot\phi}{N\sqrt z}+{\beta_2'\dot\phi}\bigg({\dot z^3\over 2N^3z^{3\over 2}}+{6k\dot z\over N\sqrt z} \bigg)-\frac{\gamma'\dot z \dot\phi}{N\sqrt z}\bigg{(}\frac{\dot z^2}{N^2z}+12k \bigg{)}+z^{\frac{3}{2}}\bigg{(}\frac{\dot{\phi}^2}{2N}-VN\bigg{)}\bigg{]}. \end{split}\ee\end{center}
Now to initiate Dirac formalism, we substitute $\dot z=Nx$, i.e.; $\ddot z=N\dot x+\dot N x$, so that the point Lagrangian may be expressed in the following form,

\begin{center}\be \label{A2}\begin{split} L = & \bigg{[}{\bigg{(}-{3\alpha'\dot\phi \sqrt z x}-\frac{3\alpha N x^2}{2\sqrt z}+6kN\alpha\sqrt z\bigg{)}}+\frac{9\beta_1}{\sqrt z}\bigg{(}\frac{\dot x^2}{N}+\frac{2kN x^2}{z}+4k^2N \bigg{)}-\frac{36\beta_1' k x \dot\phi}{\sqrt z}\\&+{\beta_2'\dot\phi}\bigg({x^3\over 2z^{3\over 2}}+{6kx\over \sqrt z} \bigg)-\frac{\gamma'x \dot\phi}{\sqrt z}\bigg{(}\frac{x^2}{z}+12k \bigg{)} +z^{\frac{3}{2}} \bigg{(}\frac{\dot{\phi}^2}{2N}-VN\bigg{)}+u\bigg({\dot z\over N}-x \bigg)\bigg{]}, \end{split}\ee\end{center}
where the expression $\big({\dot z\over N}-x \big)$ is treated as a constraint and therefore introduced through the Lagrangian multiplier $u$ in the above point Lagrangian. The canonical momenta are,

\begin{eqnarray}\begin{split}\label{Ap}
  p_x = & {18\beta_1 \dot x\over N\sqrt z}; ~~ p_z={u\over N};~~p_{\phi}=-3\alpha'\sqrt z x-{36\beta_1' kx\over \sqrt z}+\beta_2'\bigg({x^3\over 2z^{3\over 2}}+{6kx\over \sqrt z} \bigg)-{\gamma' x\over \sqrt z}{\big({x^2\over z}+12k\big)}+{\dot\phi z^{3\over 2}\over N};\\&p_N=0=p_u.
\end{split}
\end{eqnarray}
Therefore, the primary constraint Hamiltonian reads as,
\begin{equation}\label{AHp}\begin{split}
    H_{p_1}&=N\bigg{[}\frac{\sqrt z p_x^2}{36\beta_1}+\frac{p_{\phi}^2}{2z^{\frac{3}{2}}}+\frac{3\alpha' x p_{\phi}}{z}+\frac{36k\beta_1'{x} {p_{\phi}}}{z^2}-{\beta_2' p_{\phi}\over z^{3\over 2}}\bigg({x^3\over 2z^{3\over 2}}+{6kx\over \sqrt z} \bigg)+{\gamma'x p_{\phi}\over z^2}\bigg({x^2\over z}+{12k}\bigg)+\frac{9\alpha'^2 x^2}{2\sqrt z}\\& +3\alpha \bigg{(}\frac{x^2}{2\sqrt z}-2k\sqrt z\bigg{)}+\frac{108k\alpha'\beta_1' x^2}{z^{\frac{3}{2}}}-{3\alpha'\beta_2' x\over z}\bigg({x^3\over 2z^{3\over 2}}+{6kx\over \sqrt z} \bigg)+{3\alpha'\gamma'x^2\over z^{3\over 2}}\bigg({x^2\over z}+{12k}\bigg)+\frac{648k^2\beta_1'^2{x}^2}{z^{\frac{5}{2}}}\\&-\frac{18k\beta_1}{\sqrt z}\bigg{(}\frac{x^2}{z}+2k\bigg{)}-{36k\beta_1'\beta_2' x\over z^2}\bigg({x^3\over 2z^{3\over 2}}+{6kx\over \sqrt z} \bigg)+{36k\beta_1'\gamma'x^2\over z^{5\over 2}}\bigg({x^2\over z}+{12k}\bigg)+{\beta_2'^2\over 2z^{3\over 2}}\bigg({x^3\over 2z^{3\over 2}}+{6k x\over \sqrt z} \bigg)^2\\& -{\beta_2'\gamma' x\over z^2}\bigg({x^3\over 2z^{3\over 2}}+{6k x\over \sqrt z} \bigg)\bigg({x^2\over z}+12k \bigg)+{\gamma'^2x^2\over 2z^{5\over 2}}\bigg({x^2\over z}+{12k}\bigg)^2+z^{\frac{3}{2}}V\bigg{]}+\frac{u\dot{z}}{N}-u\big( {\dot z\over N}-x\big).
\end{split}
\end{equation}
Now introducing the constraints $\phi_1= Np_z-u=0$ and $\phi_2=p_u=0$ through the Lagrange multipliers $u_1$ and $u_2$ respectively, we get
\begin{equation}\label{AHp1}\begin{split}
    H_{p_1}&=N\bigg{[}\frac{\sqrt z p_x^2}{36\beta_1}+\frac{p_{\phi}^2}{2z^{\frac{3}{2}}}+\frac{3\alpha' x p_{\phi}}{z}+\frac{36k\beta_1'{x} {p_{\phi}}}{z^2}-{\beta_2' p_{\phi}\over z^{3\over 2}}\bigg({x^3\over 2z^{3\over 2}}+{6kx\over \sqrt z} \bigg)+{\gamma'x p_{\phi}\over z^2}\bigg({x^2\over z}+{12k}\bigg)+\frac{9\alpha'^2 x^2}{2\sqrt z}\\& +3\alpha \bigg{(}\frac{x^2}{2\sqrt z}-2k\sqrt z\bigg{)}+\frac{108k\alpha'\beta_1' x^2}{z^{\frac{3}{2}}}-{3\alpha'\beta_2' x\over z}\bigg({x^3\over 2z^{3\over 2}}+{6kx\over \sqrt z} \bigg)+{3\alpha'\gamma'x^2\over z^{3\over 2}}\bigg({x^2\over z}+{12k}\bigg)+\frac{648k^2\beta_1'^2{x}^2}{z^{\frac{5}{2}}}\\&-\frac{18k\beta_1}{\sqrt z}\bigg{(}\frac{x^2}{z}+2k\bigg{)}-{36k\beta_1'\beta_2' x\over z^2}\bigg({x^3\over 2z^{3\over 2}}+{6kx\over \sqrt z} \bigg)+{36k\beta_1'\gamma'x^2\over z^{5\over 2}}\bigg({x^2\over z}+{12k}\bigg)+{\beta_2'^2\over 2z^{3\over 2}}\bigg({x^3\over 2z^{3\over 2}}+{6k x\over \sqrt z} \bigg)^2\\& -{\beta_2'\gamma' x\over z^2}\bigg({x^3\over 2z^{3\over 2}}+{6k x\over \sqrt z} \bigg)\bigg({x^2\over z}+12k \bigg)+{\gamma'^2x^2\over 2z^{5\over 2}}\bigg({x^2\over z}+{12k}\bigg)^2+z^{\frac{3}{2}}V\bigg{]}+ux+u_1\big(Np_z-u\big)+u_2p_u.
\end{split}
\end{equation}
Note that the Poisson brackets $\{x, p_x\} = \{z, p_z\} = \{\phi,p_{\phi}\}=\{u, p_u\} = 1$, hold. Now constraints should remain preserved in time, which are exhibited through the following Poisson brackets
\begin{equation}\label{Apc}
   \dot\phi_1=\{ \phi_1,H_{p_1}\}=-u_2-N{\partial H_{p_1}\over \partial z}\approx 0\Rightarrow u_2=-N{{\partial H_{p_1}\over \partial z}};~ \dot\phi_2= \{\phi_2,H_{p_1}\}\approx 0\Rightarrow u_1=x.
\end{equation}
Therefore the primary Hamiltonian is modified to
\begin{equation}\label{AHp2}\begin{split}
    H_{p_2}&=N\bigg{[}xp_z+\frac{\sqrt z p_x^2}{36\beta_1}+\frac{p_{\phi}^2}{2z^{\frac{3}{2}}}+\frac{3\alpha' x p_{\phi}}{z}+\frac{36k\beta_1'{x} {p_{\phi}}}{z^2}-{\beta_2' p_{\phi}\over z^{3\over 2}}\bigg({x^3\over 2z^{3\over 2}}+{6kx\over \sqrt z} \bigg)+{\gamma'x p_{\phi}\over z^2}\bigg({x^2\over z}+{12k}\bigg)+\frac{9\alpha'^2 x^2}{2\sqrt z}\\& +3\alpha \bigg{(}\frac{x^2}{2\sqrt z}-2k\sqrt z\bigg{)}+\frac{108k\alpha'\beta_1' x^2}{z^{\frac{3}{2}}}-{3\alpha'\beta_2' x\over z}\bigg({x^3\over 2z^{3\over 2}}+{6kx\over \sqrt z} \bigg)+{3\alpha'\gamma'x^2\over z^{3\over 2}}\bigg({x^2\over z}+{12k}\bigg)+\frac{648k^2\beta_1'^2{x}^2}{z^{\frac{5}{2}}}\\&-\frac{18k\beta_1}{\sqrt z}\bigg{(}\frac{x^2}{z}+2k\bigg{)}-{36k\beta_1'\beta_2' x\over z^2}\bigg({x^3\over 2z^{3\over 2}}+{6kx\over \sqrt z} \bigg)+{36k\beta_1'\gamma'x^2\over z^{5\over 2}}\bigg({x^2\over z}+{12k}\bigg)+{\beta_2'^2\over 2z^{3\over 2}}\bigg({x^3\over 2z^{3\over 2}}+{6k x\over \sqrt z} \bigg)^2\\& -{\beta_2'\gamma' x\over z^2}\bigg({x^3\over 2z^{3\over 2}}+{6k x\over \sqrt z} \bigg)\bigg({x^2\over z}+12k \bigg)+{\gamma'^2x^2\over 2z^{5\over 2}}\bigg({x^2\over z}+{12k}\bigg)^2+z^{\frac{3}{2}}V\bigg{]}-Np_u{\partial H_{p_1}\over \partial z}.
\end{split}
\end{equation}
As the constraint should remain preserved in time in the sense of Dirac, so
\begin{equation}\label{Hp2}
    \dot\phi_1=\{\phi_1,H_{p2}\}=-N\bigg[{\partial H_{p_1}\over \partial z}-Np_u{\partial^2H_{p_1}\over \partial z^2} \bigg]+N{\partial H_{p_1}\over \partial z}
\approx 0\Rightarrow p_u=0.
\end{equation}
Finally the phase-space structure of the Hamiltonian, being free from constraints reads as,
\begin{equation}\label{AHF}\begin{split}
    H&=N\bigg{[}xp_z+ \frac{\sqrt z p_x^2}{36\beta_1}+\frac{p_{\phi}^2}{2z^{\frac{3}{2}}}+\frac{3\alpha' x p_{\phi}}{z}+\frac{36k\beta_1'{x} {p_{\phi}}}{z^2}-{\beta_2' p_{\phi}\over z^{3\over 2}}\bigg({x^3\over 2z^{3\over 2}}+{6kx\over \sqrt z} \bigg)+{\gamma'x p_{\phi}\over z^2}\bigg({x^2\over z}+{12k}\bigg)+\frac{9\alpha'^2 x^2}{2\sqrt z}\\& +3\alpha \bigg{(}\frac{x^2}{2\sqrt z}-2k\sqrt z\bigg{)}+\frac{108k\alpha'\beta_1' x^2}{z^{\frac{3}{2}}}-{3\alpha'\beta_2' x\over z}\bigg({x^3\over 2z^{3\over 2}}+{6kx\over \sqrt z} \bigg)+{3\alpha'\gamma'x^2\over z^{3\over 2}}\bigg({x^2\over z}+{12k}\bigg)+\frac{648k^2\beta_1'^2{x}^2}{z^{\frac{5}{2}}}\\&-\frac{18k\beta_1}{\sqrt z}\bigg{(}\frac{x^2}{z}+2k\bigg{)}-{36k\beta_1'\beta_2' x\over z^2}\bigg({x^3\over 2z^{3\over 2}}+{6kx\over \sqrt z} \bigg)+{36k\beta_1'\gamma'x^2\over z^{5\over 2}}\bigg({x^2\over z}+{12k}\bigg)+{\beta_2'^2\over 2z^{3\over 2}}\bigg({x^3\over 2z^{3\over 2}}+{6k x\over \sqrt z} \bigg)^2\\& -{\beta_2'\gamma' x\over z^2}\bigg({x^3\over 2z^{3\over 2}}+{6k x\over \sqrt z} \bigg)\bigg({x^2\over z}+12k \bigg)+{\gamma'^2x^2\over 2z^{5\over 2}}\bigg({x^2\over z}+{12k}\bigg)^2+z^{\frac{3}{2}}V\bigg{]}=N\mathcal{H},
\end{split}
\end{equation}
which is exactly same Hamiltonian as obtained in (\ref{Hc}). This proves that `Modified Horowitz' Formalism' is identical to `Dirac's constrained analysis', provided divergent terms are taken into account under integration by parts, prior to the initiation of constraint analysis.

\end{document}